\IfFileExists{revtex4-2.cls}{%
\documentclass[aps,prx,twocolumn,superscriptaddress,longbibliography]{revtex4-2}%
}{%
}

\usepackage{amsmath,amssymb,amsfonts,mathtools,bm}
\usepackage{empheq}
\usepackage{booktabs}
\usepackage{array}
\usepackage{enumitem}
\usepackage[hidelinks,colorlinks=true,linkcolor=blue,citecolor=blue,urlcolor=blue]{hyperref}
\allowdisplaybreaks

\newcommand{\Sym}{\operatorname{Sym}}

\newcommand{\Hilb}{\mathcal H}

\begin{document}

\title{Cluster Condensation and Symmetry-Protected Thermal Rigidity \\of Bose gases on the Haldane Sphere}

\author{Meng-Jie Wang}
\affiliation{Institute of Modern Physics, Northwest University, Xi'an 710127, China}
\author{Jun-Hui Zheng}
\email{junhui.zheng@nwu.edu.cn}
\affiliation{Institute of Modern Physics, Northwest University, Xi'an 710127, China}
\affiliation{Peng Huanwu Center for Fundamental Theory, Xi'an 710127, China}
\affiliation{Shaanxi Key Laboratory for Theoretical Physics Frontiers, Xi'an 710127, China}
\affiliation{Fundamental Discipline Research Center for  Quantum Science and technology of Shaanxi Province, Xi'an 710127, China}

\begin{abstract}
We show that a spinless Bose gas on a Haldane sphere with a small monopole
charge \(q\) realizes, in the lowest Landau level (LLL), an exactly solvable
SU(2)-protected cluster condensate: a Hafnian condensate of two-boson singlets
for \(q=1\), analogous to resonating-valence-bond singlet coverings, and a
four-boson singlet condensate for \(q=3/2\). Unlike topologically protected
quantum Hall states, the exactness of these states relies only on SU(2)
symmetry and repulsiveness, with no pseudopotential fine-tuning. Their
SU(2)-protected degeneracy renders the single-particle correlations exactly
temperature-independent even as the clusters dissociate, a thermal rigidity
absent in conventional Bose--Einstein condensates (BECs). Increasing the interaction drives a transition
to a vortex BEC with spontaneously broken rotational symmetry.
\end{abstract}

\maketitle

The Haldane sphere with a magnetic monopole at its center \cite{WuYang1976,Haldane1983} offers a powerful framework for understanding quantum Hall states \cite{Laughlin1983,HaldaneRezayi1985,Fano1986,Morf1987,Jain1989,Jain1990,Regnault2004,Greiter2011,Balram2015,Zhao2023,Moore1991,ReadRezayi1999} because of its boundary-free geometry and SU(2) rotational symmetry. 
Formerly merely a conceptual construct, this geometry has become experimentally accessible in cold-atom systems, following theoretical proposals \cite{Pietil2009,Zhou2018,Zheng2025} and successive realizations of synthetic monopoles~\cite{Ray2014,Ray2015} and bubble gases \cite{Carollo2022,Jia2022,Veyron2026}. In cold-atom experiments, the dimensionless monopole charge \(q\) is small, while the particle number and interactions between atoms remain tunable. For small \(q\), each of the \(2q\) flux quanta occupies an area comparable to the total area of the sphere, placing the system in a regime distinct from conventional quantum Hall systems. The LLL contains only \(2q+1\) single-particle states, yet in a bosonic system the many-body Hilbert space still has a large dimension, \(\binom{N+2q}{2q}\). In the absence of interactions, this Hilbert space is completely degenerate. The primary open questions are how this degeneracy is lifted by interactions and what exotic states emerge.

In this Letter, we show that in the LLL regime, where the interaction energy scale is much smaller than the Landau-level gap, a spinless Bose gas on a Haldane sphere with SU(2)-symmetric repulsive interactions, contact or long-ranged, realizes cluster BEC for \(q=1\) and \(q=3/2\). Our technical innovation is mapping the system onto an exactly solvable spin-\(q\) model, where the spin represents the orbital angular momentum of the particles. In this picture, the many-body effects reorganize the LLL Hilbert space into total-angular-momentum multiplets: different \(L\) multiplets are split, while the \((2L+1)\)-fold degeneracy within each multiplet remains protected
by SU(2) symmetry, and the ground state is the one with minimum \(L\). For \(q=1\), the ground state is a condensate of two-boson singlets, whose wavefunction takes the Hafnian form, a bosonic analog of the Resonating Valence Bond (RVB) singlet coverings \cite{Anderson1987}. For \(q=3/2\), the minimal scalar cluster contains four bosons, giving rise to a rare four-body cluster condensate. The residual SU(2) degeneracy makes the single-particle correlation exactly temperature-independent even as the clusters dissociate gradually, exhibiting {thermal rigidity}. The cluster BECs, whose degeneracy originates from SU(2) symmetry, provide a symmetry-protected counterpart to topologically protected quantum Hall states \cite{Laughlin1983,Haldane1983,Jain1989,Moore1991,ReadRezayi1999}. Crucially, this exactness does not rely on any pseudopotential fine-tuning: it is guaranteed by SU(2) symmetry and the repulsiveness of the interaction alone.

The formation of cluster singlets effectively suppresses the interaction energy by reducing particle overlap. This suggests a promising route to determining the ground state by constructing elementary cluster singlets, a powerful approach that may extend to larger \(q\). The thermal rigidity is a generic LLL feature as long as SU(2) symmetry remains unbroken, contrasting sharply with conventional BECs, where thermal
fluctuations destroy single-particle long-range order. Our results identify the Haldane sphere as a platform for cluster condensation and symmetry-protected thermal rigidity in cold-atom quantum simulation. For stronger interactions or higher temperature, the LLL approximation breaks down. At high temperature, the system becomes a normal gas from cluster BEC, while for strong interactions we confirm that it becomes a vortex BEC, in which all bosons occupy the same single-particle orbital and the rotational
symmetry is spontaneously broken.

\emph{Effective spin-\(q\) model in the LLL approximation.} We consider \(N\) bosons with mass \(M\), confined on a Haldane sphere with monopole charge \(q\in \frac{1}{2}\mathbb{Z}\) and radius \(R\). The Hamiltonian reads
\begin{equation}
\hat{H} = \hat{H}_{\rm kin} + \hat{H}_{\rm int}
= \sum_{i=1}^N \frac{\boldsymbol{\Pi}_i^2}{2M}
+ \frac{1}{2} \sum_{i\neq j}^N V(\mathbf r_i-\mathbf r_j),
\label{eq:H}
\end{equation}
where \(\boldsymbol{\Pi}=-i\hbar\boldsymbol{\nabla}'+ \mathbf A\) is the mechanical momentum with gauge potential \(\mathbf A = - \hat{\boldsymbol\phi} \,  (\hbar q/ R) \cot\theta \) \cite{Jain2007} and spherical tangential gradient \(\boldsymbol{\nabla}' =\hat{\boldsymbol{\theta}}\,R^{-1}\partial_\theta
+\hat{\boldsymbol{\phi}}\,(R\sin\theta)^{-1}\partial_\phi \). \(\hat{\boldsymbol{\theta}}\) and \(\hat{\boldsymbol{\phi}}\) are the polar and azimuthal unit vectors. The single-particle SU(2) rotation generator is
\(\hat{\boldsymbol L_i} = \mathbf r_i \times \boldsymbol{\Pi}_i + \hbar q\,{\mathbf r_i}/R \), which contains the monopole contribution and satisfies the relation, \(\boldsymbol{\Pi}_i^2=(\hat{\boldsymbol L}_i^2-\hbar^2q^2)/R^2\). The single-particle spectrum is \(E_{q,l} =[l(l+1)-q^2]\Delta_0/2\) with \(\Delta_0 ={\hbar^2}/{MR^2} \) being the characteristic kinetic energy on the sphere. The eigenfunctions are the monopole harmonics \(Y_{q,l,m}(\Omega)\), where \(\Omega=(\theta,\phi)\), \(l=|q|,|q|+1,|q|+2,\cdots\), and \(m=-l,\cdots,l\) \cite{WuYang1976}. Without loss of generality, we focus on \(q>0\);
negative \(q\) follows analogously.

We take  interactions to be SU(2)-invariant,
\([\hat{\boldsymbol{L}},\hat{H}_{\rm int}]=0\), where
\(\hat{\boldsymbol{L}}=\sum_i\hat{\boldsymbol{L}}_i\) is the total
angular-momentum operator. When the interaction energy scale is much smaller than the Landau-level gap, \(\Delta_q=E_{q,q+1}-E_{q,q}=(q+1)\Delta_0\), the system is restricted to the LLL at low temperature, \(k_B T \ll \Delta_q\) \cite{MorrisFeder2006}. In this regime, each particle has fixed angular momentum \(\hat{\boldsymbol L}_i^2 = q(q+1) \hbar^2\) and the total kinetic energy remains constant, \(E_{\rm kin}=N E_{q,q}= N q \Delta_0/2\). Thus, the low-energy physics is governed solely by the projected interaction \(\mathcal P_{\rm L}\hat H_{\rm int}\mathcal P_{\rm L}\), where \(\mathcal P_{\rm L}\) is the projector onto the LLL. For SU(2)-symmetric interactions, the total angular momentum of any two
colliding particles is conserved. This yields the Haldane pseudopotential decomposition \(\mathcal P_{\rm L} V(\mathbf r_i-\mathbf r_j) \mathcal P_{\rm L} = \sum_{J=0}^{2q}V_J\hat P_J^{ij}\) \cite{Haldane1983,Simon2007}, where \(\hat P_J^{ij}\) projects onto the interaction channel with \((\hat{\boldsymbol L}_i+\hat{\boldsymbol L}_j)^2= J(J+1)\hbar^2\). Here \(V_J \equiv \langle J,M_z|\,V(\mathbf r_i-\mathbf r_j)\,|J,M_z\rangle\), where \(|J,M_z\rangle\) is the two-boson total-angular-momentum eigenstate and \(V_J\) is independent of \(M_z\). For bosons, \(2q-J\) must be even \cite{Haldane1983}.

Below we show that for \(q\le3/2\) the projected Hamiltonian
reduces to a spin-\(q\) model, where the spin represents the orbital angular
momentum \(\hat{\boldsymbol L}_i\) of each particle. 
For \(q=1\) and \(q=3/2\), the two-boson total angular momentum is restricted to two values, \(2q\) and \(J'=2q-2 \geq 0\). The projectors onto these channels are
\begin{equation}
\hat P_{2q}^{ij}
=
\frac{(\hat{\boldsymbol L}_i+\hat{\boldsymbol L}_j)^2-J'(J'+1)\hbar^2}
{[2q(2q+1)-J'(J'+1)]\hbar^2},
\quad
\hat P_{J'}^{ij}=\hat{\mathbb{I}}-\hat P_{2q}^{ij}.
\end{equation}
Summing over all pairs \((i,j)\) gives the projected interaction [see Supplemental Material (SM) for details] \cite{zheng20262}
\begin{equation}\label{Hp}
\mathcal P_{\rm L} \hat H_{\mathrm{int}} \mathcal P_{\rm L}
=
a(q)\hat{\boldsymbol L}^{2}+b(q,N),
\end{equation}
where the coefficients are
\begin{align}
a&=
\frac{\delta V}{2(4q-1)\hbar^2}, \qquad \delta V = V_{2q} - V_{J'},
\notag\\
b&=
\frac{N(N-1)V_{J'}}{2}-
\frac{[N^2(q^2-4q+1)+(5q-1)N]\delta V}{2(4q-1)}.
\notag
\end{align}
The resulting effective spin model depends only on \(\hat{\boldsymbol L}^{2}\). Unlike conventional spin models, Bose symmetry 
requires the many-body spin state to be fully symmetric under exchange.
For \(q=1/2\), the two-boson total angular momentum must be \(J=2q\); the expression for \(\hat P_{2q}^{ij}\) still holds since \(J'(J'+1)=0\) for \(J'=-1\). Thus, Eq.\,\eqref{Hp} also holds for \(q=1/2\) with \(V_{J'}=0\).

In the SM~\cite{zheng20262}, we prove that for any repulsive interaction,
\(V_{2q}-V_{J'}>0\), and hence \(a(q)>0\). For a contact
interaction with \(\lambda>0\),
\(V(\mathbf r_i-\mathbf r_j)=\lambda\delta^{(2)}(\mathbf r_i-\mathbf r_j)\),
we obtain
\(V_{2q}={(2q+1)^2\lambda}/[{4\pi(4q+1)R^2}]\) and \(V_{J'}=0\):
only the maximal-\(J\) channel survives, since for \(J\neq 2q\) the
two-body wavefunction \(\propto(u_i v_j-u_j v_i)^{2q-J}\) vanishes
at coincidence. Here \(u_i = u(\theta_i,\phi_i)\) and \(v_i = v(\theta_i,\phi_i)\), with
\(u(\theta,\phi)=\cos(\theta/2)\,e^{i\phi/2}\) and
\(v(\theta,\phi)=\sin(\theta/2)\,e^{-i\phi/2}\) the spinor variables \cite{Haldane1983}.

Since \(a(q)>0\) for repulsive interactions, the effective model is  an antiferromagnetic spin model with coupling \(2 a(q)\sum_{i<j}\hat{\boldsymbol L}_i\cdot\hat{\boldsymbol L}_j\).  
The spectrum of Eq.~\eqref{Hp} depends only on \(L\). The allowed
values of \(L\) follow  from the irreducible decomposition of the symmetric subspace of \(N\)
spin-\(q\) particles. The ground state
minimizes \(L\), while the highest state in the LLL
maximizes it, \(L_{\max}=Nq\). The two-body interaction produces a splitting of the LLL manifold
with width \(E_s\simeq a(q)(Nq)^2\hbar^2\), which sets a
many-body interaction energy scale. Thus, the LLL approximation is
valid only when \(E_s\ll\Delta_q\). This condition ensures that the entire many-body wavefunction stays confined to the LLL subspace, and it is more stringent than the single-particle Landau-level mixing condition \cite{MorrisFeder2006}.

\emph{Exact eigen-wavefunctions and cluster condensates.}  For \(q=1/2,1,3/2\), the system realizes distinct condensates, whose
wavefunctions we construct exactly below. For \(q=1/2\), the fully symmetric subspace of \(N\) spin-\(\frac12\)
particles forms the \(L=N/2\) multiplet with degeneracy
\(\binom{N+2q}{2q}=N+1\) \cite{zheng20262}, and both the kinetic and interaction energies are constant on the LLL. This \(L=N/2\) multiplet can be spanned by the SU(2)
coherent states
\(|\theta,\phi\rangle \equiv e^{-i\phi \hat{L}_z/\hbar}\,e^{-i\theta \hat{L}_y/\hbar}|L,L\rangle\),
where \(|L,M_z=L\rangle\)
is the highest-weight state and \(|\theta,\phi\rangle\) is polarized along \((\theta,\phi)\). Each coherent state
describes a vortex BEC in which every particle occupies the same
single-particle orbital, with a vortex centered at
\((\pi-\theta,\phi+\pi)\). Explicitly,
\begin{equation}
 \Psi_{\theta,\phi}(\theta_1,\phi_1,\cdots,\theta_N,\phi_N) = \prod_{i=1}^{N}
  \left(
 u^*\,u_i + v^*\,v_i
  \right).
  \label{eq:coherent}
\end{equation}
Selecting the vortex position requires
explicit SU(2) symmetry breaking, such as a deformation of the
spherical geometry, an anisotropic interaction, or an external field
coupling to \(\hat{\boldsymbol{L}}\).  In cold-atom experiments, for
example, the gravitational potential breaks rotational symmetry.

For \(q=1\), the fully symmetric subspace of  \(N\) spin-1 particles has the irreducible decomposition \(\operatorname{Sym}(1^{\otimes N})
= N\oplus {(N-2)}\oplus {(N-4)}\oplus\cdots\oplus L_{\text{m}}\),
where the numbers denote the SU(2) spin-\(L\) irreducible representations and \(L_{\text{m}} = N\bmod2\) \cite{zheng20262}. The key principle we use to find a many-body singlet is to construct a small singlet cluster; a combination of these clusters remains a scalar singlet. We denote the LLL states by \(|m\rangle=Y_{q,q,m}\) and their annihilation operators by \(\hat a_m\), for
\(m=-q,\dots,q\). In second-quantized form, \(\hat L_z = (\hat a_1^\dagger\hat a_1 -\hat a_{-1}^\dagger\hat a_{-1})\hbar\), \(\hat L_+ = \hat L_x + i \hat L_y =  \sqrt{2} (\hat a_{1}^\dagger\hat a_0 + \hat a_{0}^\dagger\hat a_{-1})\hbar\), and \(\hat L_- = \hat L_+^\dagger\). The two-boson singlet is, up to normalization, \(|1\rangle|-1\rangle + |-1\rangle| 1\rangle - |0\rangle| 0\rangle\),  generated by \(\hat{A}^\dagger = \hat{a}_{1}^{\dagger}\hat{a}_{-1}^{\dagger} - \frac{1}{2}(\hat{a}_{0}^{\dagger})^2\), a rotational scalar satisfying \([\hat {\boldsymbol L},\hat A^\dagger]=0\) \cite{KoashiUeda2000}. Since each \(L\) appears once in the decomposition,
the highest-weight state of the spin-\(L\) multiplet is \(|L,L\rangle\propto(\hat A^\dagger)^p(\hat a_1^\dagger)^L|0\rangle\), where the \(p\) singlet pairs are scalar, while the remaining \(L\) particles in \((\hat a_1^\dagger)^L\) determine the spin-\(L\) representation, with \(N=2p+L\). The remaining components follow from the lowering operator,
\(|L, M_z\rangle\propto(\hat L_-)^{L-M_z}|L,L\rangle\). This construction generates all eigenstates of the
effective Hamiltonian \cite{zheng20262}. 

The ground
state is the \(L_{\rm m}\) multiplet. For even \(N = 2k\), it is the unique
singlet, a condensate of \(k\) two-boson singlets. In spinor variables, the unnormalized two-body singlet wavefunction is
\(D_{ij}^2\), where \(D_{ij}=u_i v_j-v_i u_j\). The
ground state is a coherent superposition of all perfect pairings, 
\begin{equation}
  |L=0,M_z=0\rangle
  \propto
  \sum_{\mathcal P}
  D_{i_1i_2}^2 D_{i_3i_4}^2 D_{i_5i_6}^2 \cdots D_{i_{2k-1}i_{2k}}^2,
\end{equation}
where \(\mathcal P\) runs over all perfect matchings of the \(N\)
particles. This is the Hafnian
\(\operatorname{Hf}[D_{ij}^2]\), the bosonic analog of an RVB singlet
covering. All pair coverings enter with the same sign, as required by
Bose symmetry, which is unlike a Pfaffian \cite{Moore1991,Regnault2004}. Because the singlet pairs are rotationally invariant, the condensate preserves SU(2) symmetry. 
Unlike the \(q=1/2\) case, where a weak anisotropy selects a direction
for the coherent state, the \(q=1\) singlet ground state is unaffected
by weak anisotropy and remains a condensate of singlet pairs. For odd \(N\), one unpaired particle remains in the \(L=1\)
multiplet, giving a threefold-degenerate ground
state.


Now we consider \(q=3/2\).  The minimum value \(L_{\min}\) depends
on \(N\bmod 4\), with \(L_{\min}=0,3/2,1,3/2\) for \(N\bmod 4=0,1,2,3\),
respectively.  Moreover, the \(L_{\min}\) multiplet appears only once
in the irreducible decomposition of
\(\operatorname{Sym}\big((3/2)^{\otimes N}\big)\) \cite{zheng20262}. We start from the construction of the smallest scalar cluster.
 Bose symmetry forbids a two-particle singlet. The smallest scalar cluster contains four bosons:
\begin{equation}
|\chi_{4s}\rangle = \mathcal S\left[D_{12}^2D_{34}^2D_{13}D_{24}\right],
\end{equation}
where \(\mathcal S\) denotes symmetrization. For each particle \(i\), the polynomial is homogeneous of degree \(3 = 2q\) in  \(u_i, v_i\), ensuring that the state resides entirely in the LLL. Acting with \(L_-=\sum_{i=1}^4 \hbar v_i\,\partial_{u_i}\) and \(L_+=\sum_{i=1}^4 \hbar u_i\,\partial_{v_i}\) \cite{Haldane1983} yields \(L_-|\chi_{4s}\rangle=L_+|\chi_{4s}\rangle=0\), confirming that it is a singlet. The quartet admits equivalent representations as two \(J=1\)
pairs or two \(J=3\) pairs coupled to zero total angular momentum,
different intermediate coupling schemes yielding the same
symmetrized four-boson state \cite{zheng20262,Varshalovich1988}.  For \(N\in 4\mathbb{Z}\),  \(L_{\min}=0\) and the multiplet appears
once. This unique \(L=0\) multiplet is a condensate of the
scalar four-boson clusters.
For general \(N\), the \(L_{\min}\) multiplet is \((2L_{\min}+1)\)-fold degenerate. The ground state is a condensate of \(\lfloor N/4 \rfloor\) scalar quartets, with the remaining \(N \bmod 4\) particles forming the \(L_{\min}\) state. The explicit construction through the four-boson scalar creation operator is presented in the SM \cite{zheng20262}.

Excitations above the \(q=3/2\) quartet condensate correspond to quartet dissociation. In contrast to \(q=1\), where breaking a singlet pair produces a unique spin-2 excited state, quartet dissociation is not unique: the fragments can carry different angular momenta, and their coupling can yield different total \(L\). Since the effective Hamiltonian depends only on \(L\) for a given \(N\), all states with the same \(L\) remain exactly degenerate, even when this \(L\) appears with multiplicity greater than one.

\emph{Thermal rigidity and cluster dissociation.} In the LLL regime, where \(E_s \ll \Delta_q\) and  \(k_B T \ll \Delta_q\), the two-point correlation function 
\begin{equation}
  G(\Omega,\Omega') = \langle\hat\Psi^\dagger(\Omega)\hat\Psi(\Omega')\rangle_T = \text{Tr}[\hat{\rho}_T \hat\Psi^\dagger(\Omega)\hat\Psi(\Omega')]
\end{equation}
exhibits thermal rigidity. Here, \(\hat{\rho}_T = e^{-\beta \hat H}/\text{Tr}[e^{-\beta \hat H}]\) is the density operator and \(\hat\Psi(\Omega) = \sum_{m} Y_{q,q,m}(\Omega)\,\hat a_m \) is the field operator. 
We first calculate \(\mathcal{G}_{nm} = \langle\hat a_m^\dagger \hat a_{n}\rangle_T\). The SU(2) symmetry implies \([\hat{\rho}_T,\hat L_a]=0\) for \(a=x\), \(y\), and \(z\),
leading to  
\(\operatorname{Tr}\!\big(\hat{\rho}_T[\hat L_a, \hat a_m^\dagger \hat a_{n}]\big)=\operatorname{Tr}\!\big([\hat{\rho}_T,\hat L_a] \hat a_m^\dagger \hat a_{n}\big)=0\). In the second-quantized form, \(\hat L_a =\sum_{km}\hat a_k^\dagger\,(L_a^{(q)})_{km}\hat a_m\), where \(L_a^{(q)}\) is the \((2q+1)\)-dimensional irreducible  representation. The Leibniz rule gives
\(\operatorname{Tr}\!\big(\hat{\rho}_T[\hat L_a, \hat a_m^\dagger \hat a_{n}]\big)=\sum_{k}(L_a^{(q)})_{km}\mathcal{G}_{nk}
-\sum_{k}(L_a^{(q)})_{nk}\mathcal{G}_{km}\). Hence,
\(\big[L_a^{(q)},\mathcal{G}\big] =0 \), and thus \(\mathcal{G} \propto \mathbb{I}\) by Schur's lemma. Since \(\sum_m\langle \hat a_m^\dagger \hat a_{m}\rangle_T=N\), we obtain
\(\langle \hat a_m^\dagger \hat a_{n}\rangle_T = [{N}/{(2q+1)}]\,\delta_{mn}\). 
This result indicates that, from a single-particle perspective, the system is
a fragmented BEC \cite{Liu2001} with all \(2q+1\) orbitals equally occupied. Using the addition theorem \cite{Fung2002}, the correlation becomes
\begin{equation}
|G(\Omega,\Omega')|
= \frac{N}{2q+1} \bigg|\sum_{m} Y^*_{q,q,m}(\Omega)Y_{q,q,m}(\Omega') \bigg|=
\frac{N}{4\pi}{\cos}^{2q}\frac{\gamma}{2},
\label{eq:G1final}
\end{equation}
exactly independent of temperature, where \(\gamma\) is the angle between \(\Omega\) and \(\Omega'\) \cite{zheng20262}. This thermal rigidity arises from SU(2) symmetry, which locks the
single-particle correlation into a fixed spatial profile, and from the
Landau-level gap \(\Delta_q\), which exponentially suppresses thermal excitation
out of the LLL. This thermal rigidity is generic to LLL bosons on the Haldane sphere,
independent of \(q\), as long as SU(2) symmetry is not spontaneously
broken. This is unlike conventional BECs, whose gapless thermal
excitations destroy single-particle long-range order.

To quantify the thermal dissociation of the clusters, we define the cluster-number operator \(\hat N_{\rm cl}\). 
For \(q=1\),  \(\hat N_{\text{cl}}\) has
eigenvalue \(p=(N-L)/2\) for the state \(|L,M_z\rangle\), and \(p\)
decreases as \(L\) increases from \(L_\text{min}\) to \(N\). Cluster formation effectively suppresses the \(V_{2q}\) channel weight,
thereby lowering the interaction energy \cite{zheng20262}.
Its
thermal average is
\(\langle \hat N_{\text{cl}}\rangle_T
= \operatorname{Tr}[\hat N_{\text{cl}} \hat{\rho}_T]\).
At zero temperature, only the ground state contributes, so
\(\langle \hat N_{\text{cl}}\rangle_T \simeq N/2\) for large \(N\).
As temperature increases, the singlets dissociate gradually.  At
sufficiently high temperature within the LLL regime, \(E_s \ll k_B T \ll \Delta_q\), all \(L\)
multiplets become equally occupied, and
\(\langle \hat N_{\text{pair}}\rangle_T \to N/6\)
\cite{zheng20262}. For \(q=3/2\), the number of states with
at least \(N_{\rm cl}\) four-body clusters is
\( F(N_{\rm cl})=
  \binom{N-4N_{\rm cl}+2q}{2q}\).
This is obtained by extracting \(N_{\rm cl}\) clusters and
distributing the remaining \(N-4N_{\rm cl}\) bosons arbitrarily among
the \(2q+1\) single-particle states.  The number of states with
exactly \(N_{\rm cl}\) clusters is then
\(d(N_{\rm cl}) = F(N_{\rm cl}) - F(N_{\rm cl}+1)\).
At zero temperature, \(\langle N_{\rm cl}\rangle_T\simeq N/4\); at
high temperature, \(\langle N_{\rm cl}\rangle_T\to N/16\), which
follows from a telescoping sum over the combinatorial multiplicities
\cite{zheng20262}.

In both cases a macroscopic number of clusters remains at all
temperatures within the LLL regime, \(k_BT\ll\Delta_q\). At higher temperatures,
when only a fraction of the particles occupy the LLL, the relevant particle number
in the above discussion should be understood as the LLL occupation
\(N_{\mathrm{L}}\), rather than the total particle number \(N\). The
above discussion then indicates that the cluster number is of the same
order as \(N_{\mathrm{L}}\). The transition from the cluster
condensate to a normal gas therefore takes place when the LLL ceases to
be macroscopically occupied. For the noninteracting case, we estimate
the BEC temperature
\(k_BT_{\mathrm{BEC}}\simeq\Delta_0 N/[2W(N/2(q+1))]\), with \(W\) the
Lambert \(W\) function, yielding
\(k_BT_{\mathrm{BEC}}\sim 10^2\Delta_0\) for \(N=10^3\)
\cite{zheng20262}.

\begin{figure}[t]
\centering
\includegraphics[width=\columnwidth]{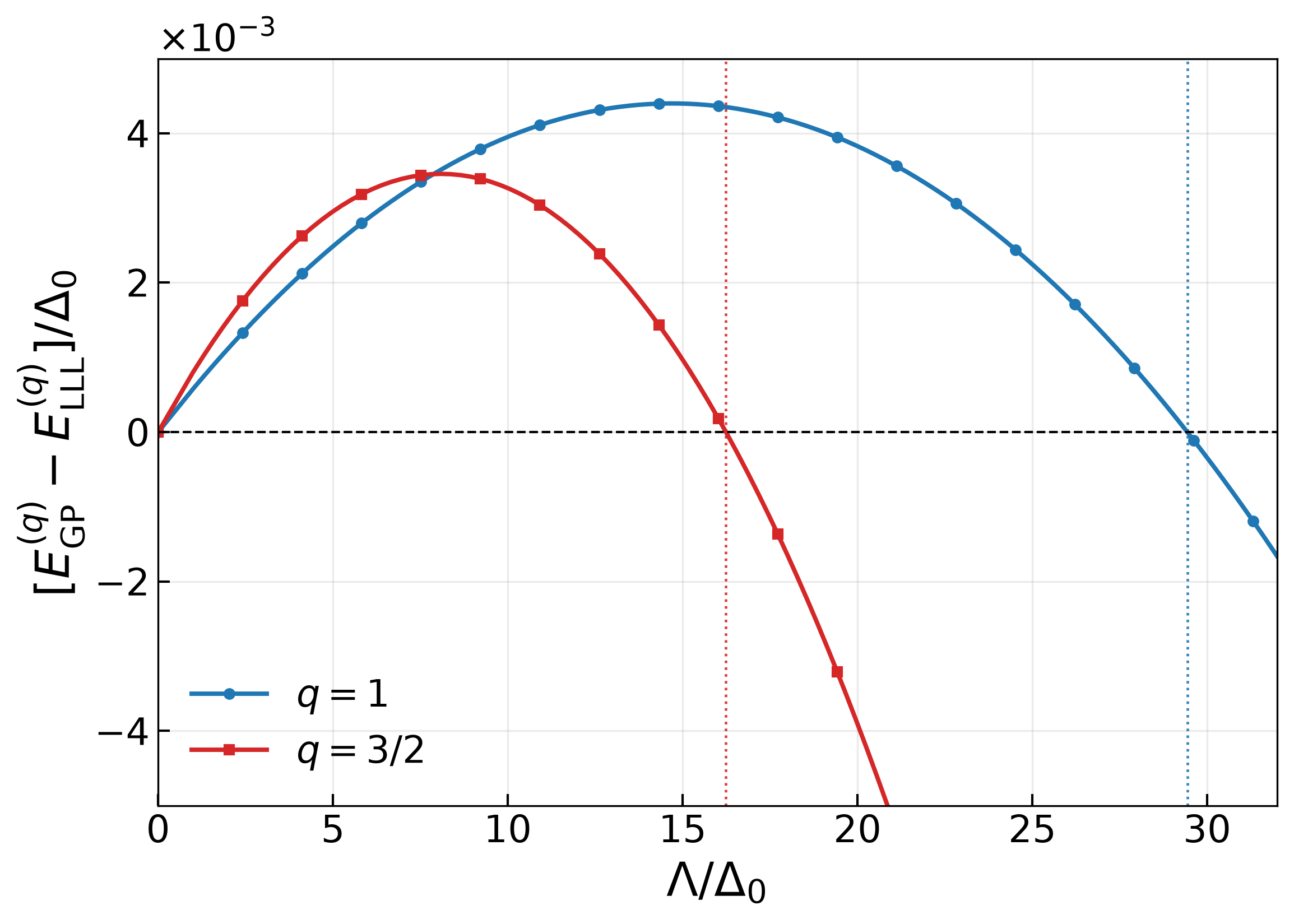}
\caption{\([E_{\rm GP}^{(q)}-E_{\rm LLL}^{(q)}]/\Delta_0\) as functions of the dimensionless interaction strength \(\Lambda/\Delta_0 \) for \(q=1,3/2\). The critical values are \(\Lambda/\Delta_0 = 29.33 \) for \(q=1\) and \(\Lambda/\Delta_0 = 16.29 \) for \(q= 3/2\) The GP solution is obtained with the cutoffs \(l_{\rm cut} = 15\). Here, \(N=10^3\).}
\label{fig:GP}
\end{figure}

{\it Breakdown of the LLL projection and the vortex BEC.} The exact solution above assumes the LLL projection. For sufficiently strong interaction, this approximation fails and the system enters a mean-field regime with suppressed density fluctuations~\cite{Cooper2001}. To examine the breakdown, we compare the LLL-projected ground-state energy with the Gross--Pitaevskii (GP) variational energy for contact interaction. The LLL-projected ground-state energy is \(E_{\rm LLL}^{(q)}= N E_{q,q}+E_{\rm int}^{(q)}(N,L_{\min})\). The GP
variational energy is obtained by minimizing 
\begin{equation}
E_{\rm GP}^{(q)} = \sum_{l,m} N E_{q,l}|c_{lm}|^2 +\frac{\lambda}{2R^2} N (N-1)\int d\Omega\,|\Phi(\Omega)|^4
\end{equation}
over \(\Phi(\Omega)=\sum_{l,m}c_{lm}Y_{q,l,m}(\Omega)\) with
\(\sum_{l,m}|c_{lm}|^2=1\).
The GP ground-state energy are converged with respect to the Landau-level cutoff \(l_{\rm cut}\). In Figure \ref{fig:GP}, we present \([E_{\rm GP}^{(q)}-E_{\rm LLL}^{(q)}]/\Delta_0\) versus \(\Lambda/\Delta_0\), where 
\({\Lambda} = \lambda N^2/4\pi R^2\) and \(N=10^3\). The energy difference changes sign at a critical \(\Lambda/\Delta_0\), signaling a transition between the cluster BEC
and the vortex BEC. At small interaction, clustering lowers the interaction energy by reducing particle overlap; at larger interaction, GP ground states become favorable due to Landau-level mixing.
The GP ground state hosts \(2q\) vortices of winding number \(1\), forming a vortex lattice ~\cite{Zhou2018,Chen2026}. For \(q=1\), two vortices sit at opposite poles, breaking SU(2) down to U(1) rotations about the vortex axis. For \(q=3/2\), three vortices are uniformly distributed along the equator, reducing the symmetry to \(\mathbb{Z}_3\) \cite{zheng20262}.

These predictions can be tested with current cold-atom technology for
synthetic monopole fields and shell-shaped geometries
\cite{Zhou2018,Zheng2025,Carollo2022}, with the interaction strength tunable via Feshbach resonances. For a \(^{87}\mathrm{Rb}\) condensate in the 
\(F=1\) hyperfine manifold \cite{Ray2014}
with shell radius \(R=0.25\,\mu\mathrm{m}\), the LLL gap is
\(\Delta_{q=1}/h\simeq3.72\,\mathrm{kHz}\), corresponding to
\(T\simeq178\,\mathrm{nK}\), well within reach of current laser-cooling
techniques. 
The comparison in Figure~\ref{fig:GP} shows that the cluster BEC gives way to
the vortex BEC at \(\Lambda/\Delta_0\simeq29.33\), corresponding to
\(E_s/\Delta_{q=1}\simeq4.4\), where the LLL approximation is no longer
valid. The gravitational potential on the shell is of order
\(MgR\simeq0.14\,\Delta_{q=1}\) per particle on Earth; a microgravity environment can suppress this effect by several orders
of magnitude.

In the GP regime, thermal excitations create vortex-antivortex pairs. At short vortex separations \(d\ll R\), their interaction reduces to the
logarithmic Coulomb form
\(\mp(2\pi\hbar^{2}n_{s}/M)\ln(d/R)\), where
\(n_{s}\simeq N/(4\pi R^{2})\) is the superfluid density, and the upper
(lower) sign corresponds to vortices with the same (opposite) winding
number \cite{Chen2026}. The system undergoes a Berezinskii–Kosterlitz–Thouless (BKT)  transition at
\(k_BT_{\mathrm{BKT}}\simeq N\Delta_0/8\) \cite{footnote}, followed by the BEC
transition at the higher temperature \(T_{\mathrm{BEC}}\)
\cite{Minnhagen1987,Nelson1977,Halperin1978,Hadzibabic2006,Tononi2019,Tononi2024}.
By contrast, in the weakly interacting LLL regime, there is no BKT
stage: the singlet dissociation is a smooth crossover, and the only
thermodynamic transition is the BEC transition,
where the LLL ceases to be macroscopically occupied.

{\it Conclusion.} 
The Haldane sphere with \(q \leq 3/2\) provides a rare exactly solvable model in the LLL. For \(q=1\) and \(3/2\), increasing the interaction drives a transition from a cluster BEC to a vortex BEC, breaking SU(2) down to U(1) or \(\mathbb{Z}_3\). Within the cluster BEC, the ground state is a Hafnian condensate of two-boson singlets for \(q=1\) and a four-boson singlet condensate for \(q=3/2\). These states are symmetry-protected
counterparts to topologically protected quantum Hall states such as the Laughlin and
Moore--Read states \cite{Laughlin1983,Moore1991};
their ground-state degeneracy arises from SU(2) symmetry rather than
topology.  The single-particle correlation is exactly
temperature-independent in the LLL regime, a thermal rigidity protected
by both the SU(2) symmetry and the Landau level gap. For \(q \geq 2\),  \(\hat{P}_{2q}\) is no longer linear in \((\hat{\boldsymbol L}_i + \hat{\boldsymbol L}_j)^2\), so whether a hidden algebraic structure survives remains open. The cluster-singlet construction offers a promising route to explore the ground-state structure for larger \(q\).

This work is supported by the National Natural Science Foundation of China under Grants No.\,12575028, and No.\,12247103.

\textbf{Supplemental Material for
``Cluster Condensation and Symmetry-Protected Thermal Rigidity  of Bose gases on the Haldane Sphere''}

This Supplemental Material provides the detailed derivations
underlying the Letter.  In Sec.~S1 we derive the effective spin-\(q\)
model from the Haldane pseudopotential decomposition, including the
projectors for \(q=1\) and \(q=3/2\), the coefficients \(a(q)\) and
\(b(q,N)\), and the pseudopotential strengths for contact and
long-ranged interactions.  In Sec.~S2 we give the angular-momentum
counting rules, including the Cayley--Sylvester formula and the irreducible decompositions of the fully symmetric subspaces for \(q=1/2\), \(1\), and \(3/2\). We also construct the exact two- and four-boson singlets, with creation
operators \(\hat A^\dagger\) and \(\hat Q^\dagger\), respectively.  In
Sec.~S3 we calculate the single-particle correlation using the
addition theorem for monopole harmonics, thereby establishing the
temperature-independent thermal rigidity discussed in the Letter.  In Sec.~S4 we
present the finite-temperature analysis of the cluster-number statistics,
including the low- and high-temperature limits of
\(\langle N_{\rm cl}\rangle_T\). We also present the
numerical results for the normalized pairing number and the thermal
quintet-channel weight for \(q=1\) with a contact interaction. In Sec.~S5 we
provide the BEC temperature estimate on the Haldane sphere (non-interacting case) and the
Lambert \(W\) function result used in the Letter.
In Sec.~S6 we present the Gross--Pitaevskii(GP) ground
state, including the full-space GP functional, the
variational expansion in monopole harmonics with 15 Landau-level
shells, and the resulting vortex configurations for \(q=1\) and
\(q=3/2\).

\section{Effective Spin Model for SU(2) Symmetric Interaction}
\label{app:spin-model}

\subsection{Effective spin model and reduction to total angular momentum}
\label{sec:effective-spin}

In this subsection we derive the effective Hamiltonian for \(N\) bosons
on a Haldane sphere in the LLL.  For \(q=1\) and \(q=3/2\), only two
two-body channels are allowed, so the projected interaction can be
expressed in terms of two-body projectors and, after summing over all
pairs, reduced to a quadratic function of the total angular momentum,
\(\hat{\boldsymbol L}^2\).  The resulting effective spin Hamiltonian is
the starting point for the exact solution and the thermal analysis
below.

In the LLL and for an SU(2)-invariant interaction, the projected 
interaction can be decomposed into Haldane pseudopotentials~\cite{sm-Haldane1983},
\begin{equation}
\mathcal{P}_{\mathrm{L}} V(\mathbf{r}_i-\mathbf{r}_j) \mathcal{P}_{\mathrm{L}}
=
\sum_{J=0}^{2q} V_J \hat{P}_J^{ij},
\end{equation}
where \(\hat{P}_J^{ij}\) projects onto the two-body channel with total angular momentum \(J\),
\begin{equation}
(\hat{\boldsymbol L}_i+\hat{\boldsymbol L}_j)^2 = J(J+1)\hbar^2,
\end{equation}
and the pseudopotential strength is
\begin{equation}
V_J \equiv \langle J,M_z|\hat{V}_{ij}|J,M_z\rangle,
\end{equation}
independent of \(M_z\). For bosons, only even \(2q-J\) channels are allowed.

For \(q=1\) and \(q=3/2\), the total angular momentum of two bosons can take only two values,
\begin{equation}
J = 2q \quad \text{and} \quad J' = 2q-2.
\end{equation}
Hence the projectors satisfy
\begin{equation}
\hat{P}_{2q}^{ij} + \hat{P}_{J'}^{ij} =  \hat{\mathbb{I}} ,
\end{equation}
and can be written explicitly as
\begin{equation}
\hat{P}_{2q}^{ij}
=
\frac{(\hat{\boldsymbol L}_i+\hat{\boldsymbol L}_j)^2 - J'(J'+1)\hbar^2}
{[2q(2q+1)-J'(J'+1)]\hbar^2},
\quad
\hat{P}_{J'}^{ij} = \hat{\mathbb{I}} - \hat{P}_{2q}^{ij}.
\label{eq:projectors}
\end{equation}
Summing over all interacting pairs, the projected interaction becomes
\begin{equation}
\mathcal{P}_{\mathrm{L}} H_{\mathrm{int}} \mathcal{P}_{\mathrm{L}}
=
\sum_{i<j} \left[
V_{2q} \hat{P}_{2q}^{ij} + V_{J'} \hat{P}_{J'}^{ij}
\right].
\end{equation}
Using \(\hat{P}_{J'}^{ij}=1-\hat{P}_{2q}^{ij}\), this can be rewritten as
\begin{equation}
\mathcal{P}_{\mathrm{L}} H_{\mathrm{int}} \mathcal{P}_{\mathrm{L}}
=
\sum_{i<j} V_{J'}
+
\left(V_{2q}-V_{J'}\right)
\sum_{i<j} \hat{P}_{2q}^{ij}.
\label{eq:Hsum}
\end{equation}
The first term is a constant:
\begin{equation}
\sum_{i<j} V_{J'} = \frac{N(N-1)}{2} V_{J'}.
\end{equation}
For the second term, we use the identity for the total angular momentum
\begin{equation}
\hat{\boldsymbol L} = \sum_{i=1}^N \hat{\boldsymbol L}_i,
\qquad
\hat{\boldsymbol L}^2 = \sum_{i=1}^N \hat{\boldsymbol L}_i^2
+ 2\sum_{i<j} \hat{\boldsymbol L}_i\cdot \hat{\boldsymbol L}_j.
\end{equation}
Since each particle in the LLL carries \(\hat{\boldsymbol L}_i^2 = q(q+1)\hbar^2\), we have
\begin{equation}
\sum_{i=1}^N \hat{\boldsymbol L}_i^2 = N q(q+1)\hbar^2.
\end{equation}
For a pair \((i,j)\),
\begin{equation}
(\hat{\boldsymbol L}_i+\hat{\boldsymbol L}_j)^2
=
\hat{\boldsymbol L}_i^2 + \hat{\boldsymbol L}_j^2
+ 2 \hat{\boldsymbol L}_i\cdot \hat{\boldsymbol L}_j.
\end{equation}
Summing over all pairs,
\begin{align}
\sum_{i<j} (\hat{\boldsymbol L}_i+\hat{\boldsymbol L}_j)^2
&=
\sum_{i<j} \left( \hat{\boldsymbol L}_i^2 + \hat{\boldsymbol L}_j^2 \right)
+ 2\sum_{i<j} \hat{\boldsymbol L}_i\cdot \hat{\boldsymbol L}_j \notag\\
&=
(N-1)\sum_{i=1}^N \hat{\boldsymbol L}_i^2
+ \left( \hat{\boldsymbol L}^2 - \sum_{i=1}^N \hat{\boldsymbol L}_i^2 \right) \notag\\
&=
\hat{\boldsymbol L}^2 + N(N-2) q(q+1)\hbar^2.
\label{eq:sumL2}
\end{align}
Substituting \eqref{eq:sumL2} into \(\sum_{i<j}\hat{P}_{2q}^{ij}\) using \eqref{eq:projectors},
\begin{equation}
\sum_{i<j} \hat{P}_{2q}^{ij}
=
\frac{
\frac{\hat{\boldsymbol L}^2}{\hbar^2} + N(N-2) q(q+1)
- \frac{N(N-1)}{2} J'(J'+1)
}{2q(2q+1)-J'(J'+1)}
.
\end{equation}
Inserting this into \eqref{eq:Hsum} yields the effective spin Hamiltonian
\begin{equation}
\mathcal{P}_{\mathrm{L}} H_{\mathrm{int}} \mathcal{P}_{\mathrm{L}}
=
a(q) \hat{\boldsymbol L}^2 + b(q,N),
\label{eq:Heff}
\end{equation}
where the coefficients are
\begin{align}
a&=
\frac{\delta V}{2(4q-1)\hbar^2},
\notag\\
b&=
\frac{N(N-1)V_{J'}}{2}-
\frac{[N^2(q^2-4q+1)+N(5q-1)]\delta V}{2(4q-1)},
\end{align}
and \(\delta V = V_{2q} - V_{J'}\).
Here we have used \(J'=2q-2\), so that
\begin{equation}
2q(2q+1)-J'(J'+1) = 8q-2.
\end{equation}
For \(q=1/2\), the two-boson total angular momentum is restricted to \(2q\) only, and the 
projector \(\hat{P}_{2q}^{ij}\) still takes the form \eqref{eq:projectors} with \(J'+1=0\). 
Equation \eqref{eq:Heff} remains valid with \(V_{J'}=0\).

\subsection{Pseudopotential strength}

In this subsection, we show that \(
  V_{2q}>V_{J'}\)
for a repulsive interaction.

We follow Haldane's construction~\cite{sm-Haldane1983}.  In the LLL on the
Haldane sphere, the two-particle wavefunction with total angular
momentum \(J\) and coherent-state parameters \((u,v)\) is
\begin{equation}
\psi^{(q,J)}_{(u,v)}(\Omega_1,\Omega_2)
\propto
(u_1v_2-u_2v_1)^{\,2q-J}
\prod_{i=1}^{2}
(u^* u_i+v^* v_i)^{J},
\label{eq:Haldane-two-particle}
\end{equation}
where \(u_i=\cos(\theta_i/2)e^{i\phi_i/2}\) and
\(v_i=\sin(\theta_i/2)e^{-i\phi_i/2}\) are the spinor coordinates of
particle \(i\), and \((u,v)\) are the coherent-state parameters,
defined analogously to \(u_i\) and \(v_i\), which specify the
orientation of the coherent state on the sphere. 
The total angular momentum takes the values \(J=0,1,\dots,2q\); for
Bose statistics, \(2q-J\) must be even.

The Haldane pseudopotential is defined by
\begin{equation}
  V_J
  =
  \int d\Omega_1\,d\Omega_2\,
  \bigl|\psi^{(q,J)}_{(u,v)}(\Omega_1,\Omega_2)\bigr|^2\,
  V(\Omega_1,\Omega_2),
  \label{eq:pseudopotential-def}
\end{equation}
where \(d\Omega=\sin\theta\,d\theta\,d\phi\).  We may shift
\(V(\Omega_1,\Omega_2)\) by an arbitrary constant,
\(V(\Omega_1,\Omega_2)\to V(\Omega_1,\Omega_2)+C\), without affecting
the difference \(V_{2q}-V_{J'}\).  Indeed, since
\(\psi^{(q,J)}_{(u,v)}(\Omega_1,\Omega_2)\) is normalized,
\begin{align}
& \int d\Omega_1\,d\Omega_2\,
\bigl|\psi^{(q,J)}_{(u,v)}(\Omega_1,\Omega_2)\bigr|^2\,
\bigl[V(\Omega_1,\Omega_2) +C\bigr] \notag\\
&= \int d\Omega_1\,d\Omega_2\,
\bigl|\psi^{(q,J)}_{(u,v)}(\Omega_1,\Omega_2)\bigr|^2\,
V(\Omega_1,\Omega_2) +C ,
\label{eq:constant-shift}
\end{align}
so the constant \(C\) cancels in the difference.  We therefore choose
the shift such that
\begin{equation}
  V(\Omega_1,\Omega_2)\ge 0.
  \label{eq:monotonic-V}
\end{equation}
A repulsive interaction means that bringing the particles closer costs
energy, i.e., \(V(\Omega_1,\Omega_2)\) decreases monotonically with the
interparticle distance.

To compare the maximal channel \(J=2q\) with the next channel
\(J'=2q-2\), we use the rotational invariance of the interaction.
For each coherent-state direction \((u,v)\), we may rotate the sphere
so that the first particle lies at the north pole.  The integral over
\(\Omega_1\) is then traded for an average over
the coherent-state direction
\(\Omega=(\theta,\phi)\).  Therefore,
\begin{equation}
  V_J
  =
  \int d\Omega\,d\Omega_2\,
  \bigl|\psi^{(q,J)}_{(u,v)}(0,\Omega_2)\bigr|^2\,
  V(0,\Omega_2).
\end{equation}
With \(u_1=1\) and \(v_1=0\), the two-particle wavefunction becomes
\begin{equation}
  \psi_{(u,v)}^{(q,J)}(0,\Omega_2)
  \propto
  (v_2)^{2q-J}\,(u^*)^{J}\,
  (u^* u_2+v^* v_2)^{J},
  \label{eq:psi-at-north-pole}
\end{equation}
where \(\Omega_2=(\theta_2,\phi_2)\) is the position of the second
particle.  Since the interaction depends only on the angular
separation, \(V(0,\Omega_2)=V(\theta_2)\).

We define the angular weight function
\begin{equation}
  f_J(\theta_2)
  =
  \int d\Omega\,
  \bigl|\psi^{(q,J)}_{(u,v)}(0,\Omega_2)\bigr|^2 ,
  \label{eq:fJ-def}
\end{equation}
where \(f_J(\theta_2)\) is normalized with respect to \(\Omega_2\),
\(\int d\Omega_2\,f_J(\theta_2)=1\).  

Expanding the coherent-state factor,
\begin{equation}
(u^* u_2+v^* v_2)^{J}
=
\sum_{m=0}^{J}
\binom{J}{m}
(u^* u_2)^{J-m}
(v^* v_2)^m ,
\label{eq:binomial-expansion}
\end{equation}
we obtain
\begin{equation}
\psi_{(u,v)}^{(q,J)}(0,\Omega_2)
\propto
\sum_{m=0}^{J}
\binom{J}{m}
(u^*)^{2J-m}(v^*)^m
u_2^{\,J-m}v_2^{\,2q-J+m}.
\label{eq:psi-expanded}
\end{equation}
Taking the modulus squared,
\begin{equation}
\begin{aligned}
&\bigl|\psi_{(u,v)}^{(q,J)}(0,\Omega_2)\bigr|^2
\propto
\sum_{m,m'=0}^{J}
\binom{J}{m}\binom{J}{m'}
(u^*)^{2J-m}(v^*)^m \\
&\quad\times(u)^{2J-m'}(v)^{m'} 
u_2^{\,J-m}v_2^{\,2q-J+m}
u_2^{*\,J-m'}v_2^{*\,2q-J+m'} .
\end{aligned}
\label{eq:psi-squared-expanded}
\end{equation}
Using the completeness relation for the coherent states,
\begin{align}
  &\int d\Omega\,
  (u^*)^{2J-m}(v^*)^m
  (u)^{2J-m'}(v)^{m'} \notag\\
  & =
  \frac{4\pi\,(2J-m)!\,m!}{(2J+1)!}\,
  \delta_{mm'} .
  \label{eq:completeness}
\end{align}
the cross terms vanish and only the diagonal terms survive.  Hence
\begin{align}
 f_J(\theta_2)  & \propto
  \sum_{m=0}^{J}
  \binom{J}{m}^2
  \frac{(2J-m)!\,m!}{(2J+1)!}
  |u_2|^{2(J-m)}
  |v_2|^{2(2q-J+m)} \notag \\
 & =
  \sum_{m=0}^{J}
  \binom{J}{m}^2
  \frac{(2J-m)!\,m!}{(2J+1)!}
  \cos^{2(J-m)}\Big(\frac{\theta_2}{2}\Big) \notag \\
  & \qquad\times
  \sin^{2(2q-J+m)}\Big(\frac{\theta_2}{2}\Big),
\end{align}
which depends only on the angular separation \(\theta_2\).

For general \(q\), it is convenient to introduce the variable
\begin{align}
   t & =\tan^2\Big(\frac{\theta_2}{2}\Big),\\
   \cos^2\Big(\frac{\theta_2}{2}\Big) & =\frac{1}{1+t},\\
   \sin^2\Big(\frac{\theta_2}{2}\Big) & =\frac{t}{1+t}.
  \label{eq:tan-variable}
\end{align}
The variable \(t\) is a monotonically increasing function of
\(\theta_2\) on \(0\le\theta_2\le\pi\), with \(t=0\) at
\(\theta_2=0\) and \(t\to\infty\) as \(\theta_2\to\pi\).  
We then obtain
\begin{equation}
  f_J(\theta_2)
  \propto
  \frac{1}{(1+t)^{2q}}
  \sum_{m=0}^{J}
  \binom{J}{m}^2
  \frac{(2J-m)!\,m!}{(2J+1)!}
  t^{\,2q-J+m}.
  \label{eq:fJ-t}
\end{equation}

We now examine how
\(f_J(\theta_2)\) varies with \(J\), focusing on the comparison
between the maximal channel \(J=2q\) and the other channel
\(J=2q-2\).

{\bf Case} \(q=1\):
Here \(J=2q=2\) and \(J'=2q-2=0\).  The two weight functions are
\begin{align}
  f_2(\theta_2)
  &\propto
  \frac{1}{(1+t)^2}
  \sum_{m=0}^{2}
  \binom{2}{m}^2
  \frac{(4-m)!\,m!}{5!}
  t^{\,m} \notag\\
  &\propto
  \frac{6+6t+t^2}{(1+t)^2},
  \label{eq:f2-q1}
\end{align}
and
\begin{equation}
  f_0(\theta_2)
  \propto
  \frac{1}{(1+t)^2}
  \binom{0}{0}^2
  \frac{0!\,0!}{1!}
  t^{\,2}
  =
  \frac{t^2}{(1+t)^2}.
  \label{eq:f0-q1}
\end{equation}
Hence the likelihood ratio is
\begin{equation}
  \frac{f_0}{f_2}
  \propto
  \frac{t^2}{6+6t+t^2}.
  \label{eq:ratio-q1}
\end{equation}
This ratio vanishes at \(t=0\), increases monotonically on \(t\ge0\), and
therefore \(f_0\) shifts probability weight towards larger \(\theta\).

{\bf Case} \(q=3/2\):
Here \(J=2q=3\) and \(J'=2q-2=1\).  The two weight functions are
\begin{align}
  f_3(\theta)
  &\propto
  \frac{1}{(1+t)^3}
  \sum_{m=0}^{3}
  \binom{3}{m}^2
  \frac{(6-m)!\,m!}{7!}
  t^{\,m} \notag\\
  &\propto
  \frac{20+30t+12t^2+t^3}{(1+t)^3},
  \label{eq:f3-q32}
\end{align}
and
\begin{equation}
  f_1(\theta)
  \propto
  \frac{1}{(1+t)^3}
  \sum_{m=0}^{1}
  \binom{1}{m}^2
  \frac{(2-m)!\,m!}{3!}
  t^{\,2+m}
  \propto
  \frac{t^2(2+t)}{(1+t)^3}.
  \label{eq:f1-q32}
\end{equation}
Hence the likelihood ratio is
\begin{equation}
  \frac{f_1}{f_3}
  \propto
  \frac{t^2(2+t)}{20+30t+12t^2+t^3}.
  \label{eq:ratio-q32}
\end{equation}
This ratio also vanishes at \(t=0\) and is monotonically increasing on
\(t\ge0\).

Since \(f_{J'}/f_{2q}\) is monotonically increasing and both weight
functions are normalized to 1, \(f_{J'}\) has less weight at small
\(\theta_2\) and more weight at large \(\theta_2\) than \(f_{2q}\).
Because \(V(\theta_2)\geq 0\) is monotonically decreasing, the average of
\(V\) with respect to \(f_{J'}\) is smaller than with respect to
\(f_{2q}\), i.e.,
\begin{equation}
  \int_0^\pi f_{J'}(\theta_2)V(\theta_2)\sin\theta_2\,d\theta_2
  <
 \int_0^\pi f_{2q}(\theta_2)V(\theta_2)\sin\theta_2\,d\theta_2.
  \label{eq:V-inequality}
\end{equation}
Therefore
\begin{equation}
  V_{2q}>V_{J'}
  \label{eq:V-inequality-boxed}
\end{equation}
for a repulsive interaction.  For the contact interaction, the same
conclusion follows directly from \(V_{J'}=0\) and \(V_{2q}>0\).

\subsection{Contact interaction}
\label{app:contact-interaction}

The contact-interacting spinless-boson problem on the sphere is a
widely used LLL model for rapidly rotating bosons 
\cite{sm-Regnault2004}. In this subsection, we calculate the corresponding
maximal-channel pseudopotential \(V_{2q}\).

In the LLL on the Haldane sphere, the single-particle state with \(m=q\) is the
monopole harmonic~\cite{sm-Jain2007}
\begin{equation}
Y_{q,q,q}(\Omega)
=
\sqrt{\frac{2q+1}{4\pi}}\,
u^{2q},
\qquad
u=\cos(\theta/2)\,e^{i\phi/2},
\label{eq:Yqqq}
\end{equation}
where the normalization constant is fixed by
\begin{equation}
\int d\Omega\,|Y_{q,q,q}(\Omega)|^2 = 1,
\qquad
d\Omega=\sin\theta\,d\theta\,d\phi .
\label{eq:normalization}
\end{equation}
The two-particle state with both bosons in the \(m=q\) orbital is
\begin{equation}
\Psi_{2q}(\Omega_1,\Omega_2)
=
Y_{q,q,q}(\Omega_1)\,
Y_{q,q,q}(\Omega_2).
\label{eq:two-particle-state}
\end{equation}
For contact interaction \(\lambda\delta^{(2)}(\mathbf r_i-\mathbf r_j)\), the Haldane pseudopotential in the maximal channel \(J=2q\) is
\begin{align}
V_{2q}
&=
\lambda
\int d\Omega_1\,d\Omega_2\,
\bigl|\Psi_{2q}(\Omega_1,\Omega_2)\bigr|^2\,
\delta^{(2)}(\mathbf{r}_1-\mathbf{r}_2),
\label{eq:V2q-def}
\end{align}
where the delta function on the sphere is
\begin{equation}
\delta^{(2)}(\mathbf{r}_1-\mathbf{r}_2)
=
\frac{\delta(\theta_1-\theta_2)\,\delta(\phi_1-\phi_2)}
{R^2\sin\theta_1}.
\label{eq:delta-sphere}
\end{equation}
Substituting \eqref{eq:two-particle-state} and \eqref{eq:delta-sphere} into
\eqref{eq:V2q-def},
\begin{equation}
\begin{aligned}
V_{2q}
&=
\frac{\lambda}{R^2}
\int d\Omega\,
\bigl|Y_{q,q,q}(\Omega)\bigr|^4 .
\end{aligned}
\label{eq:V2q-reduced}
\end{equation}
Using \eqref{eq:Yqqq},
\begin{equation}
\bigl|Y_{q,q,q}(\Omega)\bigr|^4
=
\left(\frac{2q+1}{4\pi}\right)^2
\cos^{8q}(\theta/2).
\label{eq:Y4}
\end{equation}
Therefore
\begin{equation}
V_{2q}
=
\frac{\lambda}{R^2}
\left(\frac{2q+1}{4\pi}\right)^2
\int d\Omega\,
\cos^{8q}(\theta/2).
\label{eq:V2q-integral}
\end{equation}
The angular integral is
\begin{equation}
\int d\Omega\,\cos^{8q}(\theta/2)
=
2\pi\int_0^\pi \cos^{8q}(\theta/2)\sin\theta\,d\theta
=
\frac{4\pi}{4q+1}.
\label{eq:cos-integral}
\end{equation}
Substituting \eqref{eq:cos-integral} into \eqref{eq:V2q-integral},
\begin{equation}
V_{2q}
=
\frac{(2q+1)^2\lambda}{4\pi(4q+1)R^2}.
\label{eq:V2q-final}
\end{equation}
For the contact interaction, \(V_{J'}=0\).Equivalently, the zero-range interaction acts only in the
zero-relative-angular-momentum channel, which corresponds on the
sphere to the maximal pair-angular-momentum sector \(J=2q\).
 For \(q=1\), \(J'=0\), we have
\begin{equation}
a(1)=\frac{V_{2q}}{6\hbar^2},
\qquad
b(1,N)=\frac{N(N-2)}{3}V_{2q}.
\end{equation}
Thus
\begin{equation}
E_{\mathrm{int}}^{(q=1)}(N,L)
=
\frac{V_{2q}}{6}
\left[
L(L+1)+2N(N-2)
\right].
\end{equation}
 For \(q=3/2\), \(J'=1\),
\begin{equation}
a(3/2)=\frac{V_{2q}}{10\hbar^2},
\qquad
b(3/2,N)=\frac{N(11N-26)}{40}V_{2q},
\end{equation}
so that
\begin{equation}
E_{\mathrm{int}}^{(q=3/2)}(N,L)
=
\frac{V_{2q}}{40}
\left[
4L(L+1)+11N^2-26N
\right].
\end{equation}

\section{Symmetric many-body Hilbert space and scalar cluster}
\label{sec:SM-counting}

\subsection{Irreducible decomposition}

In this subsection we decompose the many-body Hilbert space of \(N\)
bosons in the spin-\(q\) LLL into irreducible representations of the
total angular momentum.  The decomposition fixes the allowed values of
\(L\) and their multiplicities, which determine the spectrum of the
effective spin Hamiltonian. 

The orbital Hilbert space of $N$ identical bosons restricted to one spin-$q$ LLL is the completely symmetric tensor power~\cite{sm-Dolgachev2003}
\begin{equation}
\Hilb_{N,q}=\Sym^N(\Hilb_q),
\qquad
\dim\Hilb_{N,q}=\binom{N+2q}{2q}.
\label{eq:SM-symmetric-space}
\end{equation}
Its decomposition into total-angular-momentum irreducible representations is
\begin{equation}
\Sym^N(\Hilb_q)
=
\bigoplus_L\mu_{N,q}(L)\Hilb_L,
\label{eq:SM-irrep-decomposition}
\end{equation}
where $\mu_{N,q}(L)$ is the multiplicity of spin $L$.  The exact multiplicities can be found from fixed-$M_z$ occupation counting.

Define a nonnegative lowering index $j=q-m$ corresponding to the LLL state \(|m\rangle\), with $j=0,1,\ldots,2q$.  If $n_j$ particles occupy the orbital with lowering index $j$, then
\begin{equation}
\sum_{j=0}^{2q}n_j=N,
\qquad
r\equiv\sum_{j=0}^{2q}j n_j=Nq-M_z.
\label{eq:SM-lowering-count}
\end{equation}
The simultaneous generating function of particle number \(N\) and total lowering \(r\) is~\cite{sm-Dolgachev2003}
\begin{equation}
\prod_{j=0}^{2q}\frac{1}{1-tz^j}
=
\sum_{N=0}^{\infty}
\sum_{r=0}^{\infty}p_{N,q}(r)t^N z^r.
\label{eq:SM-gaussian-generating}
\end{equation}
The coefficient $p_{N,q}(r)$ counts partitions of $r$ whose Young diagrams fit inside an $N\times(2q)$ rectangle, for given $N$ and $q$. 
Therefore the dimension of the fixed-\(M_z\) subspace is
\begin{equation}
  d_{N,q}(M_z)=p_{N,q}(Nq-M_z).
  \label{eq:SM-fixed-M-dim}
\end{equation}
Every spin-\(L\) multiplet contributes one state to each magnetic
sector \(M_z=-L,-L+1,\ldots,L\).  For \(M_z\ge0\),
\begin{equation}
  d_{N,q}(M_z)=\sum_{L\ge M_z}\mu_{N,q}(L).
\end{equation}
The maximum total angular momentum is \(L_{\rm max}=Nq\), corresponding
to the highest-weight state with all particles in the \(m=q\) orbital.
For \(L>L_{\rm max}\), \(\mu_{N,q}(L)=0\), and hence
\(d_{N,q}(L_{\rm max}+1)=0\).  Subtraction of adjacent weight-space
dimensions then yields the exact Cayley--Sylvester formula
\cite{sm-Dolgachev2003}
\begin{equation}
  \mu_{N,q}(L)
  =
  d_{N,q}(L)-d_{N,q}(L+1),
  \label{eq:SM-Cayley-Sylvester}
\end{equation}
with the boundary condition
\begin{equation}
  \mu_{N,q}(L_{\rm max})=d_{N,q}(L_{\rm max}).
  \label{eq:SM-CS-boundary}
\end{equation}

To collect the multiplicities for all particle numbers,
it is convenient to define the bivariate multiplicity generating function
\begin{equation}
  \mathcal C_q(t,u)
 \equiv \sum_{N=0}^{\infty}\sum_L
     \mu_{N,q}(L)t^Nu^{2L}.
  \label{eq:SM-Molien-generating}
\end{equation}
The coefficient of $t^Nu^{2L}$ counts spin-$L$ multiplets,
whereas the specialization $\mathcal C_q(t,0)$ counts only singlets.
To obtain it directly from occupation counting, introduce
\begin{equation}
  F_q(t,u)
  \equiv \prod_{m=-q}^{q}\frac{1}{1-tu^{2m}}
  =\sum_{N,M_z}d_{N,q}(M_z)t^Nu^{2M_z}.
\end{equation}
The relation~\eqref{eq:SM-Cayley-Sylvester} gives
\begin{equation}
  \mathcal C_q(t,u)
  =\operatorname{Pos}_u\!
   \left[(1-u^{-2})F_q(t,u)\right],
\end{equation}
where $\operatorname{Pos}_u$ retains nonnegative powers of $u$.
The resulting formulas are \cite{sm-Ottaviani2013} 
\begin{eqnarray}
  \mathcal C_{\frac12}(t,u)
  &=&
  \frac{1}{1-tu}, \label{c12}\\
  \mathcal C_{1}(t,u)
  &=&
  \frac{1}{(1-tu^2)(1-t^2)}, \label{c1}\\
  \mathcal C_{\frac32}(t,u)
  &=&
  \frac{1+t^3u^3}{(1-tu^3)(1-t^2u^2)(1-t^4)}, \label{c32}
\end{eqnarray}
for \(q\leq 3/2\).

For \(q=1/2\) there are only two orbitals. The generating function in
Eq.~\eqref{eq:SM-gaussian-generating} gives
\(\prod_{j=0}^{1}(1-tz^j)^{-1}=[(1-t)(1-tz)]^{-1}\), so the fixed-\(M_z\)
dimension is \(d_{N,1/2}(M_z)=1\) for \(|M_z|\le N/2\). The
Cayley--Sylvester formula then yields
\(\mu_{N,1/2}(L)=\delta_{L,N/2}\), i.e., the completely symmetric
tensor power is a single irreducible representation,  
\begin{equation}
  \Sym^N(\Hilb_{1/2}) = \Hilb_{N/2}.
  \label{eq:SM-qhalf}
\end{equation}
Hence the only total angular momentum is \(L=N/2\).  Expanding $\mathcal C_{\frac12}(t,u)$ \eqref{c12} gives \(\sum_N t^N u^N\), i.e., \(L=N/2\), consistent with
Eq.~\eqref{eq:SM-qhalf}.

For \(q=1\), the generating
function $
  \mathcal C_{1}(t,u)$
gives
\begin{equation}
  \mu_{N,1}(L)
  =
  \begin{cases}
    1, & L=N-2b,\quad b=0,1,\dots,\lfloor N/2\rfloor,\\[2mm]
    0, & \text{otherwise},
  \end{cases}
\end{equation}
i.e., every allowed \(L=N,N-2,\dots,L_{\min}\) occurs exactly once,
with \(L_{\min}=0\) for even \(N\) and \(L_{\min}=1\) for odd \(N\).
Explicitly,
\begin{equation}
  \Sym^N(\Hilb_1)
  =
  \Hilb_N\oplus\Hilb_{N-2}\oplus\Hilb_{N-4}\oplus\cdots\oplus\Hilb_{L_{\min}}.
  \label{eq:SM-q1-decomp}
\end{equation}
As a dimension check, the left-hand side has dimension
\(\binom{N+2}{2}\), while the right-hand side gives
\begin{equation}
  \sum_{L=L_{\min},\,L_{\min}+2,\dots}^{N}(2L+1),
\end{equation}
where the sum runs over even \(L\) for even \(N\) and over odd \(L\)
for odd \(N\).  For even \(N\), this sum equals
\(\sum_{k=0}^{N/2}(4k+1)\); for odd \(N\), it equals
\(\sum_{k=0}^{(N-1)/2}(4k+3)\).  In both cases the result
equals \(\binom{N+2}{2}\), so the decomposition is complete.  This
multiplicity-free structure permits an explicit operator construction
of every many-body eigenstate in Sec.~S2.2.

For $q=3/2$, expanding the generating function $
  \mathcal C_{3/2}(t,u) $
shows that $\mu_{N,3/2}(L)$ is exactly the number of solutions of
\begin{equation}
  \begin{aligned}
    N&=a+2b+4d+3\epsilon,\\
    2L&=3a+2b+3\epsilon,
  \end{aligned}
  \label{eq:SM-q32-diophantine}
\end{equation}
where $a,b,d\ge0$ are integers and $\epsilon\in\{0,1\}$.

Minimizing $L$ at fixed $N$ gives
\begin{equation}
  L_{\min}(N)=
  \begin{cases}
    0,&N=4k,\\
    \dfrac32,&N=4k+1,\\
    1,&N=4k+2,\\
    \dfrac32,&N=4k+3,
  \end{cases}
  \qquad k=0,1,\ldots.
  \label{eq:SM-q32-Lmin}
\end{equation}
Indeed, $L=0$ requires $a=b=\epsilon=0$,
and therefore $N=4d$.
For $N=4k+2$, the smallest integer spin is $L=1$,
obtained uniquely from $b=1$ and $a=\epsilon=0$.
For odd $N$, the spin is half-integer, but $L=1/2$ is
impossible because $3a+2b+3\epsilon=1$ has no solution.
The next value, $L=3/2$, is realized uniquely by
$a=1$, $b=\epsilon=0$ for $N=4k+1$,
or by $\epsilon=1$, $a=b=0$ for $N=4k+3$.
In each case $d=k$.
Thus every lowest-spin multiplet occurs once;
in particular, $\mu_{4k,3/2}(0)=1$.
For the repulsive contact Hamiltonian derived in Sec.~S1,
the energy increases with $L(L+1)$,
so these lowest-spin multiplets are also the ground-state multiplets.

 As examples, for small \(N\) we
have 
\begin{align}
  \Sym^1(\Hilb_{\frac32}) &= \Hilb_{\frac32}, \notag\\
  \Sym^2(\Hilb_{\frac32}) &= \Hilb_3 \oplus \Hilb_1, \notag\\
  \Sym^3(\Hilb_{\frac32}) &=
  \Hilb_{\frac92} \oplus \Hilb_{\frac52} \oplus \Hilb_{\frac32}, \notag\\
  \Sym^4(\Hilb_{\frac32}) &=
  \Hilb_6\oplus\Hilb_4\oplus\Hilb_3\oplus\Hilb_2\oplus\Hilb_0, \notag \\
  \Sym^5(\Hilb_{\frac32}) &=
  \Hilb_{\frac{15}{2}}
  \oplus \Hilb_{\frac{11}{2}}
  \oplus \Hilb_{\frac92}
  \oplus \Hilb_{\frac72}
  \oplus \Hilb_{\frac52}
  \oplus \Hilb_{\frac32}, \notag \\
  \Sym^6(\Hilb_{\frac32}) &=
  \Hilb_9\oplus\Hilb_7\oplus\Hilb_6\oplus\Hilb_5
  \oplus\Hilb_4\oplus2\Hilb_3\oplus\Hilb_1.
\notag
\end{align}
The repeated \(L=3\) representation at \(N=6\) shows that multiplicity
labels do occur.  Nevertheless, the effective spin Hamiltonian depends
only on \(N\) and \(L\), so all copies of the same \(L\) remain exactly
degenerate.

\subsection{Scalar cluster and multiplet construction for \(q=1\)}\label{ss}

Introduce creation operators \(\hat a_m^\dagger\) for the three LLL
orbitals \(m=1,0,-1\) and define the scalar pair creator
\begin{equation}
  \hat A^\dagger
  =
  \hat a_1^\dagger\hat a_{-1}^\dagger
  -\frac12(\hat a_0^\dagger)^2,
  \label{eq:SM-A-dagger}
\end{equation}
which satisfies
\begin{align}
  [\hat L_z,\hat A^\dagger] &= 0, \qquad
  [\hat L_\pm,\hat A^\dagger] = 0, \\
  [\hat N,\hat A^\dagger] &= 2\hat A^\dagger, \qquad
  [\hat A,\hat A^\dagger] = \hat N+\frac32.
\end{align}
It is also proportional to the standard rank-zero tensor obtained by
coupling two spin-$1$ creation operators
\cite{sm-KoashiUeda2000,sm-Varshalovich1988},
\begin{equation}
[\hat{\bm a}^\dagger\times\hat{\bm a}^\dagger]^{(0)}_0
= 
  \frac{1}{\sqrt3}
  \sum_{m=-1}^{1}
  (-1)^{m}\,
  \hat a_m^\dagger\hat a_{-m}^\dagger =
-\frac{2}{\sqrt3}\hat A^\dagger.
\end{equation}

For fixed \(N\) and an allowed \(L\), let
\begin{equation}
  p=\frac{N-L}{2}.
  \label{eq:SM-p-number}
\end{equation}
The state \((\hat a_1^\dagger)^L|0\rangle\) is the highest-weight seed
of spin \(L\).  Multiplying by the scalar \((\hat A^\dagger)^p\) adds
\(2p\) particles without changing \(L\).  Hence
\begin{equation}
  |L,M_z =L\rangle
  \propto
  (\hat A^\dagger)^p(\hat a_1^\dagger)^L|0\rangle
  \label{eq:SM-q1-highest-un}
\end{equation}
is a nonzero highest-weight state in the unique spin-\(L\)
representation.  Because Eq.~\eqref{eq:SM-q1-decomp} is
multiplicity-free, this construction identifies the exact eigenspace.

Take the highest-weight seed
\begin{equation}
  |v\rangle = (\hat a_1^\dagger)^L|0\rangle ,
  \label{eq:SM-seed}
\end{equation}
which has particle number \(v=L\) and satisfies
\(\hat A|v\rangle=0\), since \(|v\rangle\) contains only \(m=1\) particles.

Using \([\hat A,\hat A^\dagger]=\hat N+\frac32\), one proves by
induction that
\begin{equation}
  \hat A(\hat A^\dagger)^p|v\rangle
  =
  p\left(p+L+\frac12\right)(\hat A^\dagger)^{p-1}|v\rangle
  \label{eq:SM-A-through-p}
\end{equation}
for \(p\ge1\).  For \(p=1\), the commutator gives
\(\hat A\hat A^\dagger|v\rangle=(L+\frac32)|v\rangle\), which agrees
with the right-hand side.  Assuming the relation holds for \(p\), we
write
\begin{equation}
  \hat A(\hat A^\dagger)^{p+1}|v\rangle
  =
  [\hat A,(\hat A^\dagger)^{p+1}]|v\rangle
  +
  (\hat A^\dagger)^{p+1}\hat A|v\rangle .
\end{equation}
The second term vanishes because \(\hat A|v\rangle=0\).  Using
\begin{equation}
  [\hat A,(\hat A^\dagger)^{p+1}]
  =
  \sum_{k=0}^{p}(\hat A^\dagger)^k
  \left(\hat N+\frac32\right)
  (\hat A^\dagger)^{p-k} ,
\end{equation}
and \(\hat N(\hat A^\dagger)^{p-k}|v\rangle
=(L+2(p-k))(\hat A^\dagger)^{p-k}|v\rangle\), we obtain
\begin{equation}
  [\hat A,(\hat A^\dagger)^{p+1}]|v\rangle
  =
  (p+1)\left(p+1+L+\frac12\right)(\hat A^\dagger)^p|v\rangle ,
\end{equation}
which completes the induction.

The squared norm then follows from
Eq.~\eqref{eq:SM-A-through-p},
\begin{equation}
  \mathcal N_{N,L}^2
  =
  \|(\hat A^\dagger)^p(\hat a_1^\dagger)^L|0\rangle\|^2
  =
  L!\,p!\,
  \frac{\Gamma(p+L+3/2)}{\Gamma(L+3/2)} ,
  \label{eq:SM-q1-norm}
\end{equation}
with \(p=(N-L)/2\).  Hence the normalized highest-weight state is
\begin{equation}
  |L,L\rangle
  =
  \frac{(\hat A^\dagger)^{(N-L)/2}(\hat a_1^\dagger)^L}
  {\sqrt{L![(N-L)/2]!}}
  \left[
  \frac{\Gamma(L+3/2)}{\Gamma((N+L+3)/2)}
  \right]^{1/2}
  |0\rangle .
  \label{eq:SM-q1-highest-normalized}
\end{equation}
All other magnetic components are generated by lowering,
\begin{equation}
  |L,M_z\rangle
  =
  \left[
  \frac{(L+M_z)!}{(2L)!(L-M_z)!}
  \right]^{1/2}
  \left(\frac{\hat L_-}{\hbar}\right)^{L-M_z}
  |L,L\rangle .
  \label{eq:SM-q1-allM}
\end{equation}
The integer \(p=(N-L)/2\) is therefore an exact collective
scalar-pair number for this operator construction.  It does not mean
that the indistinguishable bosons are partitioned into fixed
classical dimers.

It is convenient to rescale the pair creator by
\begin{equation}
  \hat\Theta^\dagger
  =
  2\hat A^\dagger
  =
  2\hat a_1^\dagger\hat a_{-1}^\dagger
  -
  (\hat a_0^\dagger)^2 .
  \label{eq:SM-Theta}
\end{equation}
The factor of \(2\) is chosen so that the normalized ground state for
even \(N\) takes a particularly simple form.  Starting from the
normalized state in Eq.~\eqref{eq:SM-q1-highest-normalized} with
\(L=0\) and \(p=N/2\), and using
\((\hat A^\dagger)^{N/2}=2^{-N/2}(\hat\Theta^\dagger)^{N/2}\), we
obtain
\begin{equation}
  |\Psi_{g}\rangle
  =
  \frac{1}{\sqrt{(N+1)!}}
  (\hat\Theta^\dagger)^{N/2}|0\rangle .
  \label{eq:SM-Theta-state}
\end{equation}
To verify the normalization, we use
\(\hat\Theta^\dagger=2\hat A^\dagger\) and the squared norm
\begin{equation}
  \|(\hat A^\dagger)^{M}|0\rangle\|^2
  =
  M!\,
  \frac{\Gamma(M+3/2)}{\Gamma(3/2)}
  =
  M!\,
  \frac{(2M+1)!!}{2^M} ,
\end{equation}
with \(M=N/2\).  Hence
\begin{align}
  \|(\hat\Theta^\dagger)^{M}|0\rangle\|^2
 & =
  4^M
  \|(\hat A^\dagger)^{M}|0\rangle\|^2
  =
  2^M M!\,(2M+1)!! \notag\\
 & =
  (2M+1)!
  =
  (N+1)! ,
\end{align}
so that
\(\langle\Psi_g|\Psi_g\rangle=1\).


\subsection{Scalar-cluster for \(q=3/2\)}

For \(q=3/2\) and \(N=4\), the full symmetric space has dimension
\begin{equation}
  \dim\Sym^4(\Hilb_{3/2})=\binom{7}{4}=35.
\end{equation}
Moreover,
\begin{equation}
  \Sym^4(\Hilb_{3/2})
  =
  \Hilb_6\oplus\Hilb_4\oplus\Hilb_3\oplus\Hilb_2\oplus\Hilb_0.
\end{equation}
Thus the four-boson scalar is unique up to normalization and overall
phase.  In the occupation convention
\begin{equation}
  (n_{3/2},n_{1/2},n_{-1/2},n_{-3/2}),
\end{equation}
the \(M_z=0\) basis contains five states,
\begin{equation}
  \begin{split}
    &|0,2,2,0\rangle,
    \quad |0,3,0,1\rangle,
    \quad |1,0,3,0\rangle,\\
    &|1,1,1,1\rangle,
    \quad |2,0,0,2\rangle.
  \end{split}
  \label{eq:SM-q32-M0-basis}
\end{equation}
The most general quartic \(M=0\) creation polynomial is a linear
combination of the five monomials associated with
Eq.~\eqref{eq:SM-q32-M0-basis}.  Imposing the highest-weight condition
\([\hat L_+,\hat Q^\dagger]=0\) leaves one free overall scale, where
\(\hat Q^\dagger\) is the creation operator of the four-boson scalar.
In second-quantized form,
\begin{align}
  \hat L_z &=
  \hbar\sum_m m\,\hat a_m^\dagger\hat a_m, \\
  \hat L_+ &=
  \hbar\sum_m
  \sqrt{\left(\tfrac32-m\right)\left(\tfrac52+m\right)}\,
  \hat a_{m+1}^\dagger\hat a_m, \\
  \hat L_- &=
  \hbar\sum_m
  \sqrt{\left(\tfrac32+m\right)\left(\tfrac52-m\right)}\,
  \hat a_{m-1}^\dagger\hat a_m.
\end{align}
The condition \([\hat L_+,\hat Q^\dagger]=0\) then fixes the relative
coefficients, and we obtain
\begin{equation}
  \begin{aligned}
    \hat Q^\dagger
    ={}&
    (\hat a_{1/2}^\dagger)^2(\hat a_{-1/2}^\dagger)^2
    -\frac{4}{\sqrt3}
    \left[
    (\hat a_{1/2}^\dagger)^3\hat a_{-3/2}^\dagger
    +\hat a_{3/2}^\dagger(\hat a_{-1/2}^\dagger)^3
    \right]
    \\
    &+6\hat a_{3/2}^\dagger\hat a_{1/2}^\dagger
    \hat a_{-1/2}^\dagger\hat a_{-3/2}^\dagger
    -3(\hat a_{3/2}^\dagger)^2(\hat a_{-3/2}^\dagger)^2.
  \end{aligned}
  \label{eq:SM-Qdagger}
\end{equation}
The condition
\([\hat L_z,\hat Q^\dagger]=0\) is automatically satisfied because
\(\hat Q^\dagger\) is built from \(M_z=0\) monomials.
Its action on the vacuum is
\begin{align}
  \hat Q^\dagger|0\rangle
  ={}&
  2|0,2,2,0\rangle
  -4\sqrt2|0,3,0,1\rangle
  -4\sqrt2|1,0,3,0\rangle
  \nonumber\\
  &+6|1,1,1,1\rangle
  -6|2,0,0,2\rangle.
\end{align}
The normalized four-boson singlet is
\begin{align}
  |\chi_{4s}\rangle
  &=\frac{1}{\sqrt{140}}\hat Q^\dagger|0\rangle \notag\\
  & =\frac{1}{\sqrt{35}}
  \bigl[
  |0,2,2,0\rangle
  -2\sqrt2|0,3,0,1\rangle -2\sqrt2|1,0,3,0\rangle \notag\\
  & ~~~~
  +3|1,1,1,1\rangle
  -3|2,0,0,2\rangle
  \bigr].
  \label{eq:SM-q32-normalized}
\end{align}
The operator has the exact quantum numbers
\begin{equation}
  [\hat N,\hat Q^\dagger]=4\hat Q^\dagger,
  \qquad
  [\hat L_z,\hat Q^\dagger]
  =[\hat L_\pm,\hat Q^\dagger]=0.
  \label{eq:SM-Q-scalar}
\end{equation}
It adds four bosons without changing the total angular momentum
of the parent state.

The explicit polynomial in Eq.~\eqref{eq:SM-Qdagger} has a simple
irreducible-tensor interpretation \cite{sm-Varshalovich1988}.  Let
\(\hat P_{1,m}^\dagger\) create a normalized symmetric two-boson state
in the allowed \(J=1\) channel.  A convenient choice is
\begin{align}
  \hat P_{1,1}^\dagger
  &=\frac{1}{\sqrt5}
  \left(\sqrt3\,\hat a_{3/2}^\dagger\hat a_{-1/2}^\dagger
  -(\hat a_{1/2}^\dagger)^2\right),\\
  \hat P_{1,0}^\dagger
  &=\frac{1}{\sqrt{10}}
  \left(3\hat a_{3/2}^\dagger\hat a_{-3/2}^\dagger
  -\hat a_{1/2}^\dagger\hat a_{-1/2}^\dagger\right),\\
  \hat P_{1,-1}^\dagger
  &=\frac{1}{\sqrt5}
  \left(\sqrt3\,\hat a_{1/2}^\dagger\hat a_{-3/2}^\dagger
  -(\hat a_{-1/2}^\dagger)^2\right).
\end{align}
Coupling two such rank-one tensors to total rank zero gives
\begin{equation}
  \left[\hat P_1^\dagger\times\hat P_1^\dagger\right]^{(0)}_0 =\frac{1}{\sqrt{3}}\sum_{m=-1}^{1}
(-1)^m
\hat P_{1,m}^\dagger\hat P_{1,-m}^\dagger 
  =-\frac{\sqrt3}{10}\hat Q^\dagger.
  \label{eq:SM-Q-coupled}
\end{equation}
The quartet is therefore not an \emph{ad hoc} linear combination; it
is the standard scalar obtained by recoupling two symmetric \(J=1\)
pairs.

Note that the four-boson singlet admits different intermediate
constructions.  In particular, it can also be obtained as the scalar
formed by recoupling two symmetric \(J=3\) pairs.  Let
\(\hat P_{3,M}^\dagger\) create a normalized symmetric two-boson state
in the allowed \(J=3\) channel,
\begin{align}
  \hat P_{3,3}^\dagger
  &=\frac{1}{\sqrt2}(\hat a_{3/2}^\dagger)^2,
  \nonumber\\
  \hat P_{3,2}^\dagger
  &=\hat a_{3/2}^\dagger\hat a_{1/2}^\dagger,
  \nonumber\\
  \hat P_{3,1}^\dagger
  &=\frac{1}{\sqrt{10}}
  \left(
    2\hat a_{3/2}^\dagger\hat a_{-1/2}^\dagger
    +\sqrt3(\hat a_{1/2}^\dagger)^2
  \right),
  \nonumber\\
  \hat P_{3,0}^\dagger
  &=\frac{1}{\sqrt{10}}
  \left(
    \hat a_{3/2}^\dagger\hat a_{-3/2}^\dagger
    +3\hat a_{1/2}^\dagger\hat a_{-1/2}^\dagger
  \right),
  \nonumber\\
  \hat P_{3,-1}^\dagger
  &=\frac{1}{\sqrt{10}}
  \left(
    2\hat a_{1/2}^\dagger\hat a_{-3/2}^\dagger
    +\sqrt3(\hat a_{-1/2}^\dagger)^2
  \right),
  \nonumber\\
  \hat P_{3,-2}^\dagger
  &=\hat a_{-1/2}^\dagger\hat a_{-3/2}^\dagger,
  \nonumber\\
  \hat P_{3,-3}^\dagger
  &=\frac{1}{\sqrt2}(\hat a_{-3/2}^\dagger)^2 .
  \label{eq:SM-P3-components}
\end{align}
Substitution of Eq.~\eqref{eq:SM-P3-components} gives the unnormalized
scalar contraction
\begin{equation}
  \sum_{M=-3}^{3}(-1)^M
  \hat P_{3,M}^\dagger\hat P_{3,-M}^\dagger
  =\frac{3}{10}\hat Q^\dagger .
  \label{eq:SM-P3-unnormalized-contraction}
\end{equation}
Thus the \(J=1\) and \(J=3\) recoupling schemes do not generate two
independent four-boson singlets.  Since the scalar sector of
\(\operatorname{Sym}^{4}(\mathcal H_{3/2})\) is one-dimensional, both
nonzero constructions must be proportional to the same operator
\(\hat Q^\dagger\).

In first-quantized spinor coordinates the Letter uses
\begin{equation}
  |\chi_{4s}\rangle
  \propto
  \mathcal S\left[D_{12}^{2}D_{34}^{2}D_{13}D_{24}\right],
  \label{eq:SM-chi4s}
\end{equation}
where \(\mathcal S\) symmetrizes the particle labels. Note that for each particle \(i\), the polynomial is homogeneous of degree \(3 = 2q\) in the spinor variables \(u_i\) and \(v_i\), which ensures that the state fully resides in the LLL.  This state is
symmetric, nonzero, and obeys \(L_\pm|\chi_{4s}\rangle=0\), hence has
\(L=0\).  Because we have proved that the symmetric four-boson scalar subspace is one-dimensional, Eq.~\eqref{eq:SM-chi4s} and Eq.~\eqref{eq:SM-q32-normalized} necessarily represent the same physical state up to normalization and an overall phase.

For \(N=4k\) and \(L=0\), since the scalar subspace is one-dimensional,
\begin{equation}
  |L=0, M_z=0\rangle
  \propto
  (\hat Q^\dagger)^k|0\rangle
  \label{eq:SM-many-quartet}
\end{equation}
is the exact ground state for repulsive contact interaction.  This is
the precise algebraic content of the scalar-quartet
condensate in the Letter.  For \(N=4k+r\) with \(r=1,2,3\), the same
scalar background leaves a unique minimal residual angular momentum
\(L_{\min}=3/2,1,3/2\), respectively.

\section{Calculate the single-particle correlation using the addition theorem}
\label{app:thermal-rigidity}

The single-particle
correlation gives
\begin{equation}
  G(\Omega,\Omega')
  =
  \frac{N}{2q+1}
  \sum_{m=-q}^{q}
  Y_{q,q,m}^*(\Omega)
  Y_{q,q,m}(\Omega') .
  \label{eq:app-G-addition}
\end{equation}
The sum is the addition theorem for monopole harmonics with \(l=q\)~\cite{sm-Jain2007}.
To evaluate it explicitly, we use the spinor representation
\begin{equation}
  Y_{q,q,m}(\Omega)
  =
  \sqrt{\frac{2q+1}{4\pi}\binom{2q}{q+m}}
  \,u^{q+m}v^{q-m}.
\end{equation}
Hence
\begin{align}
  &\sum_{m=-q}^{q}
  Y_{q,q,m}^*(\Omega)
  Y_{q,q,m}(\Omega')
  \nonumber\\
  &\quad=
  \frac{2q+1}{4\pi}
  \sum_{m=-q}^{q}
  \binom{2q}{q+m}
  (u^* u')^{q+m}(v^*v')^{q-m} \nonumber
  \\
  &\quad=
  \frac{2q+1}{4\pi}
  (u^*u'+v^*v')^{2q} .
\end{align}
Here, we have used the binomial theorem~\cite{sm-Fung2002}.

It remains to evaluate \(Z\equiv u^*u'+v^*v'\). 
Its squared modulus is
\begin{align}
  |Z|^2
  &=
  \cos^2\frac{\theta}{2}\cos^2\frac{\theta'}{2}
  +
  \sin^2\frac{\theta}{2}\sin^2\frac{\theta'}{2}
  \nonumber\\
  &\quad+
  2\cos\frac{\theta}{2}\cos\frac{\theta'}{2}
  \sin\frac{\theta}{2}\sin\frac{\theta'}{2}
  \cos\Delta\phi,
\end{align}
where 
\(\Delta\phi=\phi'-\phi\).
Using the half-angle identities, the first two terms combine to
\begin{equation}
  \cos^2\frac{\theta}{2}\cos^2\frac{\theta'}{2}
  +
  \sin^2\frac{\theta}{2}\sin^2\frac{\theta'}{2}
  =
  \frac{1}{2}\left(1+\cos\theta\cos\theta'\right),
\end{equation}
while the cross term is
\begin{equation}
  2\cos\frac{\theta}{2}\cos\frac{\theta'}{2}
  \sin\frac{\theta}{2}\sin\frac{\theta'}{2}
  \cos\Delta\phi
  =
  \frac{1}{2}\sin\theta\sin\theta'\cos\Delta\phi .
\end{equation}
Hence
\begin{equation}
  |Z|^2
  =
  \frac{1}{2}(1+\cos\gamma)
  =
  \cos^2\frac{\gamma}{2} ,
\end{equation}
where \(\gamma\) is the angle between \(\Omega\) and \(\Omega'\),
\begin{equation}
  \cos\gamma
  =
  \cos\theta\cos\theta'
  +
  \sin\theta\sin\theta'\cos(\phi-\phi') .
\end{equation}
Since \(\gamma\in[0,\pi]\), we have \(\cos(\gamma/2)\ge0\), and
therefore
\begin{equation}
  |Z|=\cos\frac{\gamma}{2}.
\end{equation}
As a result,
\begin{equation}
  {
  |G_1(\gamma)|
  =
  \frac{N}{4\pi}
  \cos^{2q}\frac{\gamma}{2}.
  }
  \label{eq:app-G1-final}
\end{equation}

\section{Thermally averaged cluster number}
\label{sec:general-probe}

For \(q=1\), at high temperature all states become equally populated, so that
\begin{equation}
  \langle N_{\rm cl}\rangle_T
  \to
  \frac{\sum_L (2L+1)\,p(L)}{\sum_L (2L+1)} ,
\end{equation}
where \(p(L)=(N-L)/2\) and the sum runs over the allowed values
\(L=L_{\min},L_{\min}+2,\dots,N\).

The denominator is the dimension of the symmetric space,
\begin{equation}
  \sum_L(2L+1)=\dim\Sym^N(\Hilb_1)=\binom{N+2}{2}
  \simeq\frac{N^2}{2}.
\end{equation}
For the numerator, writing \(L=2k\) for even \(N\),
\begin{equation}
  \sum_L(2L+1)\frac{N-L}{2}
  =
  \frac12\left[
  N\binom{N+2}{2}
  -
  \sum_L L(2L+1)
  \right].
\end{equation}
The second sum is dominated by
\begin{equation}
  \sum_L L(2L+1)
  \simeq
  2\sum_{k=0}^{N/2}(2k)^2
  \simeq
  \frac{N^3}{3},
\end{equation}
so that
\begin{equation}
  \sum_L(2L+1)\frac{N-L}{2}
  \simeq
  \frac12\left[
  \frac{N^3}{2}-\frac{N^3}{3}
  \right]
  =
  \frac{N^3}{12}.
\end{equation}
Hence
\begin{equation}
  \langle N_{\rm cl}\rangle_T
  \to
  \frac{N^3/12}{N^2/2}
  =
  \frac{N}{6}.
\end{equation}

For \(q=3/2\), at high temperature all configurations are equally populated, so
\begin{equation}
  \langle N_{\rm cl}\rangle_T
  \to
  \frac{S}{Z},
  \qquad
  Z=\sum_{k=0}^{K}d(k),\quad
  S=\sum_{k=0}^{K}k\,d(k),
\end{equation}
with \(K=\lfloor N/4\rfloor\) and
\begin{equation}
  d(k)=F(k)-F(k+1),\qquad
  F(k)=\binom{N-4k+3}{3}.
\end{equation}
The telescoping sums give
\begin{equation}
  Z=F(0)-F(K+1)=\binom{N+3}{3},
\end{equation}
and
\begin{equation}
  S=\sum_{k=1}^{\lfloor N/4\rfloor}F(k)
  =
  \sum_{j\in\{N-4,N-8,\cdots\}}\binom{j+3}{3}.
\end{equation}
Approximating the sum by an integral with step \(4\),
\begin{equation}
  S\simeq\frac14\int_0^N\binom{j+3}{3}dj
  \simeq
  \frac14\cdot\frac{N^4}{24}
  =
  \frac{N^4}{96}.
\end{equation}
Therefore
\begin{equation}
  \langle N_{\rm cl}\rangle_T
  \to
  \frac{N^4/96}{\binom{N+3}{3}}
  \simeq
  \frac{N^4/96}{N^3/6}
  =
  \frac{N}{16}.
\end{equation}

To probe the thermal dissociation of the cluster, we consider the
two-body and four-body total-angular-momentum operators
\(\hat{\boldsymbol L}_{ij}=\hat{\boldsymbol L}_i+\hat{\boldsymbol L}_j\)
and
\(\hat{\boldsymbol L}_{ijkl}=\hat{\boldsymbol L}_i+\hat{\boldsymbol L}_j+\hat{\boldsymbol L}_k+\hat{\boldsymbol L}_l\).
For \(q=1\) the relevant observable is
\(\langle \hat{\boldsymbol L}_{ij}^2\rangle_T\), while for \(q=3/2\)
it is \(\langle \hat{\boldsymbol L}_{ijkl}^2\rangle_T\).  An increase in
either expectation value signals a decrease in the number of cluster
singlets.  For four particles, the square expands as
\begin{equation}
  \hat{\boldsymbol L}_{ijkl}^2
  =
  \sum_{a\in\{i,j,k,l\}} \hat L_a^2
  +
  2\sum_{a<b} \hat{\boldsymbol L}_a\cdot\hat{\boldsymbol L}_b .
  \label{eq:L4expand}
\end{equation}
Using
\begin{equation}
  2\hat{\boldsymbol L}_a\cdot\hat{\boldsymbol L}_b
  =
  (\hat{\boldsymbol L}_a+\hat{\boldsymbol L}_b)^2
  -
  \hat L_a^2-\hat L_b^2 ,
  \label{eq:dot2L2}
\end{equation}
Eq.~\eqref{eq:L4expand} becomes
\begin{equation}
  \hat{\boldsymbol L}_{ijkl}^2
  =
  \sum_{a<b\in\{i,j,k,l\}}
  (\hat{\boldsymbol L}_a+\hat{\boldsymbol L}_b)^2
  + \text{const.}
  \label{eq:L4decomp}
\end{equation}
The two quantities
are linearly related, so that the dissociation is captured by two-particle correlation function
\(g_2(T)=\langle(\hat{\boldsymbol L}_i+\hat{\boldsymbol L}_j)^2\rangle_T\).
Since \(g_2(T)=\sum_{J=2q,J'}\langle \hat P^{12}_J\rangle_T J(J+1)\hbar^2\), dissociation corresponds to increased occupation of the maximal \(J=2q\) channel and thus to an increase in the interaction energy.

\begin{figure}[t]
\centering
\includegraphics[width=\columnwidth]{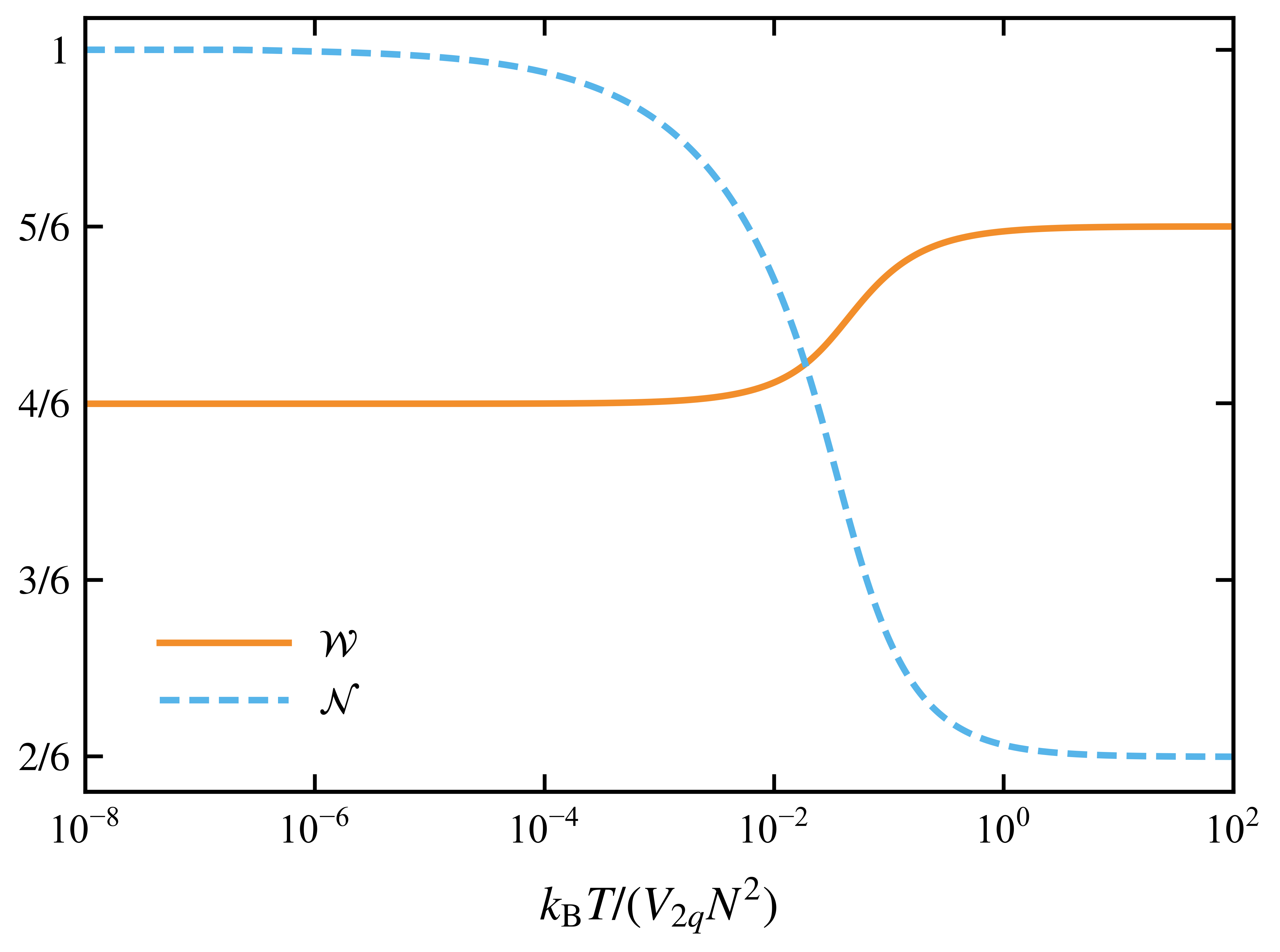}
\caption{The normalized pairing number
\(\mathcal{N} = {2\langle \hat N_{\text{pair}} \rangle_T}/{N} \) and the  quintet-channel weight 
\(\mathcal{W}  = \langle \hat H_{\text{int}}\rangle_T/[ V_{2q} C(N)] \) as functions of the dimensionless temperature \(k_{\rm B}T/( V_{2q} N^2)\), where \(C(N) = N(N-1)/2\) is the total number of pairs. Here, \(q=1\) and \(N=10^{3}\).}
\label{fig1}
\end{figure}

We take \(q=1\) with a contact interaction as an example.We introduce the normalized pairing number \( \mathcal{N} = {2\langle \hat N_{\text{cl}} \rangle_T}/{N} \). At zero temperature, only the ground state contributes, so \(\mathcal{N} \to 1\) for \(N\gg 1\). As temperature increases, higher-\(L\) multiplets become thermally occupied, and \(\mathcal{N} \to 1/3\). Since only the \(J=2q\) channel contributes to the interaction energy, we define \(C(L) = E_{\text{int}}/( V_{2q})\), which measures how many pairs of particles interact through the quintet \((J=2)\) channel for a given eigenstate. The highest-weight  state \(L=N\) has \(C(L=N) = N(N-1)/2\), with all pairs interacting through the quintet channel. For the ground state, \(C/C(N) \to 2/3\) for \(N\gg 1\), meaning about \(2/3\) of the pairs interact through this channel. The thermal quintet-channel weight is \(\mathcal{W} = \langle \hat H_{\text{int}}\rangle_T/[ V_{2q} C(N)]\), which increases from \(2/3\) to \(5/6\) with temperature. Since \(\hat H_{\rm int}\) is quartic in the field operators, \(\mathcal W\) is an integrated normally ordered second-order, or equivalently four-point, correlation. Its temperature dependence therefore shows that the thermal rigidity of \(G\) does not extend to higher-order correlations. The numerical results for \(\mathcal{N}\) and \(\mathcal{W}\) are plotted in Figure~\ref{fig1}. The monotonic decrease and increase, respectively, show that thermal fluctuations progressively suppress singlet pairing. The smoothness of these curves, with no singularity at any temperature, indicates that this dissociation of singlets occurs without a phase transition. The finite-temperature LLL treatment requires \(k_B T\ll\Delta_q\), which suppresses the real thermal occupation of higher Landau levels. 

\section{Estimating $T_{\text{BEC}}$}

In this section we estimate the noninteracting BEC temperature on
the Haldane sphere and compare it with the LLL gap \(\Delta_q\).

Here \(T_{\rm BEC}\) is used as an ideal-gas finite-size reference
scale.  Bose--Einstein condensation is characterized by a macroscopic
eigenvalue of the one-body reduced density matrix in the
Penrose--Onsager sense \cite{sm-PenroseOnsager1956}.  If several
eigenvalues are simultaneously macroscopic, the condensate is
fragmented.  Since the LLL on the sphere has
degeneracy \(2q+1\), the saturation criterion below diagnoses
macroscopic occupation of the LLL manifold and does not by itself
imply occupation of a unique single-particle orbital.

The single-particle Landau levels on the sphere are labelled by
\cite{sm-Greiter2011}
\begin{equation}
  l=q+n,\qquad n=0,1,2,\dots,
\end{equation}
with energy
\begin{equation}
  E_{q,l}
  =
  \frac{\Delta_0}{2}\left[l(l+1)-q^2\right].
\end{equation}
The degeneracy of the \(l\)-th level is \(2l+1\).  The LLL is
\(l=q\), with energy \(E_{q,q}=q\Delta_0/2\).  The excitation energy
relative to the LLL is
\begin{equation}
  E_{q,l}-E_{q,q}
  =
  \frac{\Delta_0}{2}\left[l(l+1)-q(q+1)\right].
\end{equation}
The gap to the first excited level \(l=q+1\) is
\begin{equation}
  \Delta_q
  =
  E_{q,q+1}-E_{q,q}
  =
  (q+1)\Delta_0 .
\end{equation}

We define \(T_{\rm BEC}\) as the temperature at which the excited
levels can just accommodate all \(N\) particles when the chemical
potential reaches the LLL energy from below,
\(\mu\to E_{q,q}^-\).  The saturated population of the excited levels
is then
\begin{equation}
  N_{\rm ex}^{\rm sat}(T)
  =
  \sum_{l=q+1}^{\infty}
  \frac{2l+1}
  {\exp\left[\frac{l(l+1)-q(q+1)}{2t}\right]-1},
  \qquad
  t\equiv\frac{k_BT}{\Delta_0}.
  \label{eq:app-Nex}
\end{equation}
The BEC temperature \(t_{\rm BEC}\) is fixed by
\(N_{\rm ex}^{\rm sat}(T_{\rm BEC})=N\), i.e.,
\begin{equation}
  N
  =
  \sum_{l=q+1}^{\infty}
  \frac{2l+1}
  {\exp\left[\frac{l(l+1)-q(q+1)}{2t_{\rm BEC}}\right]-1}.
  \label{eq:app-TBEC-discrete}
\end{equation}
For \(q=1\) and \(q=3/2\), this reads
\begin{align}
  q=1:&\quad
  N=\sum_{n=1}^{\infty}\frac{2n+3}
  {\exp\left[\frac{n(n+3)}{2t_{\rm BEC}}\right]-1},
  \label{eq:app-q1-sum}\\
  q=3/2:&\quad
  N=\sum_{n=1}^{\infty}\frac{2n+4}
  {\exp\left[\frac{n(n+4)}{2t_{\rm BEC}}\right]-1}.
  \label{eq:app-q32-sum}
\end{align}

\begin{figure*}[t]
\centering
\includegraphics[width=1.8\columnwidth]{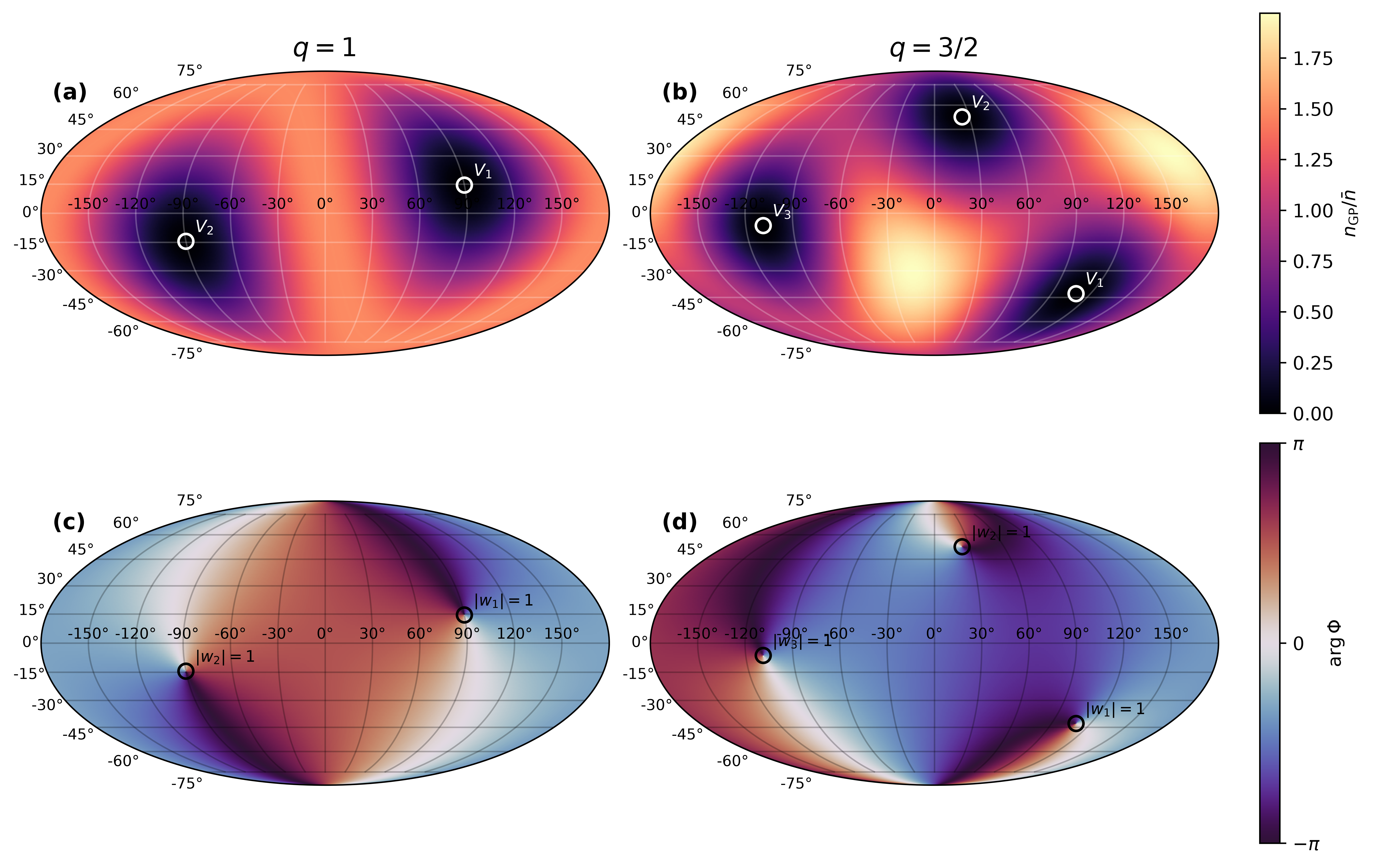}
\caption{Normalized density (a,b) and phase (c,d) on the sphere obtained from
the GP solution for \(N=10^3\), computed with a cutoff
\(l_{\rm cut}=15\) at \(\Lambda/\Delta_0=35\).  (a) and (c) are for
\(q=1\), where two vortices appear at opposite poles.  (b) and (d)
are for \(q=3/2\), where three vortices are uniformly distributed
 along a great circle.}
\label{figsm2}
\end{figure*}

For \(N\gg1\), many levels are occupied and the sum can be
approximated by an integral (semiclassical approximation).  Writing \(l\) as a continuous variable,
\begin{equation}
  \frac{dE}{dl}=\frac{\Delta_0}{2}(2l+1),
\end{equation}
so that the density of states is constant,
\begin{equation}
  \rho(E)=\frac{2}{\Delta_0}.
\end{equation}
The lower limit of \(E\) is \(\Delta_q=(q+1)\Delta_0\).  Hence
\begin{align}
  N_{\rm ex}^{\rm sat,sc}(T)
  & =
  \frac{2}{\Delta_0}
  \int_{\Delta_q}^{\infty}
  \frac{dE}{e^{E/k_BT}-1} \notag \\
  &=
  \frac{2k_BT}{\Delta_0}
  \ln\left[
  \frac{1}{1-e^{-\Delta_q/k_BT}}
  \right].
\end{align}
Setting \(N_{\rm ex}^{\rm sat,sc}(T_{\rm BEC})=N\) gives
\begin{equation}
  N
  =
  2t_{\rm BEC}^{\rm sc}
  \ln\left[
  \frac{1}{1-e^{-(q+1)/t_{\rm BEC}^{\rm sc}}}
  \right].
  \label{eq:app-TBEC-sc}
\end{equation}
For \(N\gg1\) one has \(t_{\rm BEC}^{\rm sc}\gg q+1\), so that
\begin{equation}
  1-e^{-(q+1)/t_{\rm BEC}^{\rm sc}}
  \simeq
  \frac{q+1}{t_{\rm BEC}^{\rm sc}},
\end{equation}
and Eq.~\eqref{eq:app-TBEC-sc} reduces to
\begin{equation}
  N
  \simeq
  2t_{\rm BEC}^{\rm sc}
  \ln\left(\frac{t_{\rm BEC}^{\rm sc}}{q+1}\right).
  \label{eq:app-TBEC-log}
\end{equation}
Using the Lambert \(W\) function, \(W(x)e^{W(x)}=x\), the solution of
Eq.~\eqref{eq:app-TBEC-log} is
\begin{equation}
  t_{\rm BEC}^{\rm sc}
  \simeq
  \frac{N}{2W\!\left[\frac{N}{2(q+1)}\right]}.
\end{equation}
Restoring units,
\begin{equation}
{
  k_BT_{\rm BEC}
  \simeq
  \frac{\Delta_0 N}
  {2W\!\left[\frac{N}{2(q+1)}\right]}.
  }
  \label{eq:app-TBEC-Lambert}
\end{equation}

For \(N=10^3\), solving Eq.~\eqref{eq:app-TBEC-discrete} numerically
gives
\begin{align}
  q=1:&\quad
  k_BT_{\rm BEC}\simeq106.66\,\Delta_0, \\
  q=3/2:&\quad
  k_BT_{\rm BEC}\simeq111.52\,\Delta_0 .
\end{align}
The semiclassical formula~\eqref{eq:app-TBEC-Lambert} gives
\(121.70\,\Delta_0\) and \(127.23\,\Delta_0\), respectively, in good
agreement.  Since
\begin{equation}
  \frac{k_BT_{\rm BEC}}{\Delta_q}
  =
  \frac{k_BT_{\rm BEC}}{(q+1)\Delta_0}
  \sim
  \mathcal O(N/\ln N)
  \gg
  1,
\end{equation}
the BEC temperature lies far above the LLL gap.

For a \(^{87}\)Rb condensate with \(R=0.25\,\mu\text{m}\), one finds
\(\Delta_{q=1}/h\simeq3.72\,\text{kHz}\), corresponding to
\(T\simeq178\,\text{nK}\).  For \(N=10^3\), the BEC temperature is of order
\(10\,\mu\text{K}\), and the LLL regime is reached at
\(T\ll178\,\text{nK}\), which is well within the reach of current
laser-cooling techniques.

\section{GP ground state}
\label{sec:SM-GP}

Beyond the LLL projection, the symmetry-broken mean-field branch is
obtained by minimizing the full-space Gross--Pitaevskii functional \cite{sm-Cooper2008}
\begin{equation}
E_{\rm GP}^{(q)}[\Phi]
=
N\langle\Phi|\hat h_0|\Phi\rangle
+
\frac{\lambda N(N-1)}{2R^2}
\int d\Omega\,|\Phi(\Omega)|^4, \\
\end{equation}
where
\begin{equation}
\int d\Omega\,|\Phi|^2=1. \label{eq:SM-GP-functional}
\end{equation}
Here, the factor \(R^{-2}\) converts the solid-angle measure
\(d\Omega\) into the physical area measure.  The eigenstates of
\(\hat h_0\) are the monopole harmonics \(Y_{q,l,m}\), with
\begin{equation}
E_{q,l}
=
\frac{\Delta_0}{2}
\left[l(l+1)-q^2\right],
\label{eq:SM-GP-spectrum}
\end{equation}
Denote \(
l=q+n\),
the variational orbital is expanded as
\begin{equation}
\Phi(\Omega)
=
\sum_{n=0}^{14}
\sum_{m=-q-n}^{q+n}
c_{nm}Y_{q,q+n,m}(\Omega),
\qquad
\sum_{n,m}|c_{nm}|^2=1,
\label{eq:SM-GP-expansion}
\end{equation}
and the coefficients \(c_{nm}\) are determined by numerical
minimization of Eq.~\eqref{eq:SM-GP-functional}.  We retain 15
Landau-level shells, \(n=0,\ldots,14\), corresponding to
\(l_{\max}=15\) for \(q=1\) and \(l_{\max}=31/2\) for \(q=3/2\), at which both the GP energy and vortex configuration are well
converged.
The interaction strength is parametrized by
\begin{equation}
\frac{\Lambda}{\Delta_0}
=
\frac{\lambda N^2}{4\pi R^2\Delta_0}.
\label{eq:SM-GP-interaction}
\end{equation}

Figure~\ref{figsm2} shows the normalized density and phase of the
optimized GP orbital for \(N=10^3\) at
\(\Lambda/\Delta_0=35\), on the GP-favored side of the variational
energy comparison. Since
\begin{equation}
n_{\rm GP}(\Omega)=\frac{N}{R^2}|\Phi(\Omega)|^2,
\qquad
\bar n=\frac{N}{4\pi R^2},
\end{equation}
the density displayed in the upper panels is
\(n_{\rm GP}/\bar n=4\pi|\Phi|^2\).

Each density zero coincides with a local \(2\pi\) phase winding and
therefore represents a singly quantized vortex rather than an ordinary
density depletion.  For \(q=1\), the two vortices occupy antipodal
points, producing an axially symmetric configuration about the line
joining their cores.  For \(q=3/2\), three vortices are equally spaced
along a great circle and form a configuration with discrete threefold
symmetry.  In both cases, the vortex number counted with multiplicity
is \(2q\), as fixed by the monopole flux.  Because the Hamiltonian is
rotationally invariant, the absolute orientation selected by the
numerical minimization is arbitrary: all globally rotated
configurations are energetically degenerate, whereas the vortex
number, winding, and relative arrangement are physically meaningful.

\end{document}